\documentclass[pre,nofootinbib,floatfix,superscriptaddress, twocolumn]{revtex4}
\usepackage{dcolumn}
\usepackage{graphicx}
\usepackage{amsmath,amsfonts,amssymb, epsfig}
\usepackage{amsthm}
\usepackage{natbib}
\usepackage{epstopdf}
\usepackage{verbatim}
\usepackage[caption=false]{subfig}
\usepackage{color}
\usepackage{rotating}
\usepackage[pdftex,bookmarks, colorlinks, citecolor =blue, breaklinks]{hyperref}

\def\XXint#1#2#3{{\setbox0=\hbox{$#1{#2#3}{\int}$}
\vcenter{\hbox{$#2#3$}}\kern-.5\wd0}}

\renewcommand{\Re}{{\mathrm{Re}}}
\renewcommand{\Im}{{\mathrm{Im}}}
\newcommand{\diag}{{\mathrm{diag}}}
\newcommand{\Eq}[1]{(\ref{eq:#1})}
\newcommand{\Sec}[1]{\S\ref{sec:#1}}
\newcommand{\br}{{\bf r}}
\newcommand{\bu}{{\bf u}}
\newcommand{\vphi}{{\varphi}}
\newcommand{\eps}{{\varepsilon}}

\newtheorem*{assumption*}{\assumptionnumber}
\providecommand{\assumptionnumber}{}
\makeatletter
\newenvironment{assumption}[2]
 {%
  \renewcommand{\assumptionnumber}{Asmp.~#1$\mathcal{#2}$}%
  \begin{assumption*}%
  \protected@edef\@currentlabel{#1$\mathcal{#2}$}%
 }
 {%
  \end{assumption*}
 }
\makeatother

\newcommand{\Asmp}[1]{Asmp.~\ref{assum:#1}}

\begin{document}
\title{The Hamiltonian Mean  Field model: effect of network structure on synchronization dynamics}
\author{Yogesh S. Virkar}
\email{Yogesh.Virkar@colorado.edu}
\affiliation{Department of Computer Science, University of Colorado at Boulder, Boulder, CO, 80309, USA}
\author{Juan G. Restrepo}
\email{Juan.Restrepo@colorado.edu}
\author{James D. Meiss}
\email{James.Meiss@colorado.edu}
\affiliation{Department of Applied Mathematics, University of Colorado at Boulder, Boulder, CO 80309-0526, USA}

\begin{abstract}
The Hamiltonian Mean Field (HMF) model of coupled inertial, Hamiltonian rotors  is a prototype for conservative dynamics in systems with long-range interactions. 
We consider the case where the interactions between the rotors are governed by a network described by a weighted adjacency matrix. By studying the linear stability of the incoherent state, we find that the transition to synchrony occurs at a coupling constant $K$ inversely proportional to the largest eigenvalue of the adjacency matrix. We derive a closed system of equations for a set of local order parameters and use these equations to study the effect of network heterogeneity on the synchronization of the rotors. We find that for values of  $K$ just beyond the transition to synchronization the degree of synchronization is highly dependent on the network's heterogeneity, but that for large values of $K$ the degree of synchronization is robust to changes in the heterogeneity of the network's degree distribution. Our results are illustrated with numerical simulations on Erd\"os-Renyi networks and networks with power-law degree distributions.
\end{abstract}

\keywords{network synchronization, complex networks, hamiltonian mean field model}

\maketitle

\section{Introduction}
\label{sec:introduction}

The Hamiltonian Mean Field (HMF) model \cite{barre:et:al:2006, campa:giansanti:morelli:2007, dauxois:et:al:2002, antoni:ruffo:1995, konishi:kaneko:1992, inagaki:konishi:1993} is a paradigmatic model for conservative systems exhibiting long-range interactions. Examples of such systems include free electron lasers \cite{antoniazzi:et:al:2006}, rarefied plasmas \cite{dauxois:et:al:2002}, gravitational $n$-body problems \cite{chavanis:vatteville:bouchet:2005}, etc. This model has attracted attention due to its striking dynamical properties which include second order phase transitions and violent relaxation towards persistent meta-equilibrium states \cite{lyndenbell:1967}.

The generalized HMF model describes the dynamics of $N$ interacting rotors with phase angles and angular momenta $\{(\theta_n, p_n): n=1, 2, \dots N\}$ through the Hamiltonian 
\begin{align} \label{eq:hamiltonianNetworkCase}
	H = \frac{1}{2} \sum_{n=1}^{N} \frac{p_n^2}{I_n} - \frac{K}{2N} \sum_{n, m=1}^N A_{nm} 
		\cos\left(\theta_m - \theta_n\right) \enspace.
\end{align} 
Here the first sum represents the kinetic energy of the rotors with moments of inertia $I_n$, and the second the potential energy  of coupling through a network adjacency matrix $A$ where $A_{nm} \neq 0$ if there is an edge from node $m$ to node $n$ and $A_{nm} = 0$ otherwise. For $A_{nm} > 0$, the potential energy due to the interaction of rotors $n$ and $m$ is minimized when they are aligned, $\theta_n = \theta_m$. Without loss of generality $A$ can be taken to be symmetric, $A^T = A$, since the asymmetric part of $A$ does not contribute to the interaction term in \Eq{hamiltonianNetworkCase}. The overall coupling strength is represented by $K$ and it is scaled by $1/N$ so that the energy per rotor has a finite limit as $N \to \infty$.

While the HMF model has been proposed as a model for systems with long-range interactions, in its commonly studied form these interactions are assumed to be of such long range that the rotors are all-to-all coupled [$A_{nm} \equiv 1$ in (1)]. A natural question is what is the effect of allowing a more general form for the interaction network. Such a generalization would include spatially distributed systems with decaying interactions, varying interaction strengths, and arbitrary interaction structure. Since the case of heterogenous moments of inertia was considered by two of the authors in \cite{restrepo:meiss:2014}, we will assume  in the current paper that the moments of inertia are identical, $I_n \equiv 1$,  focusing on the effects of the network structure on the dynamics.
Most previous studies for the HMF model except \cite{chavanis:vatteville:bouchet:2005, restrepo:meiss:2014, ciani:fanelli:ruffo:2011, nigris:leoncini:2013} have considered the all-to-all case with $A_{nm} \equiv 1$ in \Eq{hamiltonianNetworkCase}. Chavanis et al. \cite{chavanis:vatteville:bouchet:2005} consider stellar (gravitational) systems with interactions depending on the mass $M = I$ of the stars and thus $A_{nm} = M_{n} M_{m}$. Restrepo and Meiss \cite{restrepo:meiss:2014} study the disordered HMF model where $A_{nm} = a_n a_m$, and  $a$ and $I$ have independent, heterogeneous distributions. In terms of the network structure, both of these variants of the HMF model can be thought of as dynamics on a weighted, all-to-all network. Ciani, Fanelli, and Ruffo \cite{ciani:fanelli:ruffo:2011} studied the HMF model on Erd\"os-Renyi networks. 
Another generative model, the Watt-Strogatz small-world network \cite{watts:strogatz:1998},  was used by Nigris and Leoncini \cite{nigris:leoncini:2013}. Both \cite{nigris:leoncini:2013} and \cite{barabasi:albert:1999} obtain a description of the dynamics in terms of the network model parameters that requires a model-fitting step. In contrast to these previous approaches that study specific network ensembles, in this paper we will develop a more general theory that applies to any given network described by its adjacency matrix $A$. To test our theory, we will use Erd\"os-Renyi networks and networks with heterogenous degree distributions, such as networks with power-law degree distributions, i.e., ``scale-free" networks \cite{barabasi:albert:1999}. While we will use both Erd\"os-Renyi and scale-free networks in our examples, we emphasize that our analysis does not rely on an assumed generative mechanism for the network: it works directly with the network adjacency matrix. 

Below we determine onset of instability of the incoherent state, obtain a self-consistent equation for a set of local order parameters and quantify the degree of synchrony in terms of a macroscopic global order parameter $R$. As in previous studies on network synchronization, e.g. \cite{restrepo:ott:hunt:2005}, we find that the principal eigenvalue $\lambda$ of the network adjacency matrix is a key quantity in determining the onset of synchronization. Finally, we quantify the maximum achievable synchrony for a given network structure and find that this maximum value is very robust to the heterogeneity of the network's degree distribution. 

The rest of the paper is organized as follows. The model and its governing dynamical equations are described in \Sec{HMFOnNetwork}. In \Sec{linearStabilityAnalysis} we discuss the linear stability of the incoherent solution. We use this analysis to find the critical value of the coupling constant for the onset of synchronization.
We then study the synchronized state in \Sec{synchronizedState} and give results for the global order parameter as a function of the coupling strength in terms of a set of self-consistent equations for the local order parameters. We provide approximations to the solution of these equations just past the onset of synchrony and in the strong coupling limit. Finally, we discuss our results in \Sec{conclusion}.

\section{Network HMF Model}
\label{sec:HMFOnNetwork}

In the original HMF model, and in most subsequent studies \cite{barre:et:al:2006, campa:giansanti:morelli:2007, dauxois:et:al:2002, antoni:ruffo:1995}, all rotors in \Eq{hamiltonianNetworkCase} were assumed to have the same moments of inertia, $I_n \equiv 1$, and the coupling was assumed to be all-to-all with equal strength, $A_{mn} \equiv 1$.
While such a simplified setting provides many insights, interactions are rarely uniform and all-to-all in practice. For example, the HMF model is a simplified model for an $n$-body gravitational system in one spatial dimension with periodic boundary conditions, keeping only one harmonic of the potential \cite{inagaki:konishi:1993,chavanis:vatteville:bouchet:2005}; in this case, the interaction strength should be proportional to the product of the particle masses and decay with the separation of the particles.

With this motivation we allow for a general adjacency matrix, $A$, in \Eq{hamiltonianNetworkCase}, but simplify by setting $I_n \equiv 1$.
The resulting dynamical system is 
\begin{align} 
	\dot{\theta}_n &= p_n \label{eq:dPhase_dT} \enspace,\\
	\dot{p}_n &= \frac{K}{N} \sum_{m=1}^{N} A_{nm} \sin\left(\theta_m - \theta_n \right) \label{eq:dMomentum_dT} \enspace.
\end{align}
As is usual, it is convenient to define order parameters to quantify synchronization. When the network is heterogeneous, one can define a set of real, local order and phase parameters, $\{(R_n, \psi_n): n = 1, \ldots N\}$,  by
\begin{align}\label{eq:localOrderParam}
	R_n e^{i\psi_n} = \frac{1}{N} \sum_{m=1}^{N} A_{nm} e^{i\theta_m} \enspace,
\end{align}
that characterize the coherence of inputs to a given node.
Using these, \Eq{dMomentum_dT} becomes
\begin{align} \label{eq:dMomentum_dT_orderParam}
	\dot{p}_n = K R_n \sin\left(\psi_n - \theta_n\right)
\end{align}
The overall synchrony of rotors can be measured by a global order parameter \cite{restrepo:ott:hunt:2005}
\begin{align} \label{eq:globalOrderParam}
R = \frac{1}{\| d \|}\sum_{n=1}^{N} R_n \enspace.
\end{align}
Here $\| \dots \|$ denotes the average over nodes,
\begin{equation}\label{eq:NodeAverage}
	\| X \| \equiv \frac{1}{N} \sum_{n=1}^{N} X_n \enspace,
\end{equation}
 and $d_n$ denotes the effective degree of the $n^{th}$ node, 
\[
 	d_n \equiv \sum_{m=1}^{N} A_{nm} \enspace. 
\]
The normalization in \Eq{globalOrderParam} is chosen so that $R=1$ if all rotors are in synchrony ($\theta_n = \theta_m$).

\section{Linear stability analysis}
\label{sec:linearStabilityAnalysis}
In this section we study the incoherent state, in which the local order parameters $R_n$ are approximately zero and the rotors evolve approximately independently of each other, i.e., \eqref{eq:dPhase_dT} and \eqref{eq:dMomentum_dT} become
$\dot{\theta}_n = p_n$, $\dot{p}_n = 0 $. 
In this case $\theta_n(t) = p_n(0) t + \theta_n(0)$. Assuming that the initial momenta
differ, $p_n(0) \neq p_m(0)$ for $m \neq n$, then each oscillator has a different frequency and \eqref{eq:localOrderParam} gives $\langle |R_n|^2 \rangle_t = \sum_{m=1}^{N} A_{nm}^2/N^2$.
Here $\langle \ldots \rangle_t$ denotes a time average,
\begin{equation}\label{eq:timeAverage}
	\langle X \rangle_t \equiv \frac{1}{T_2-T_1} \int_{T_1}^{T_2} X(t) dt.
\end{equation}
In general we will choose an initial time $T_1$ large enough to eliminate transient behavior, and the interval $T_2-T_1$ large enough to reduce fluctuations.
In order that the order parameters be small, we require that $\sum_{m=1}^{N} A_{nm}^2 \ll N^2$. In particular, in our examples we have $A_{nm}^2 = A_{nm}$ and so this condition becomes $d_n \ll N^2$. Under this assumption $R_n = 0$ for all $n$ is an approximate solution of the system. In this section, we will study the stability of this incoherent state using a method similar to that used in Refs.~\cite{daido:1990, restrepo:ott:hunt:2005}.

\subsection{Dispersion relation and onset of instability}
\label{sec:dispersion}

\begin{table*}[htb]
\centering{
\scalebox{0.95}{
   \begin{tabular}{|c|c|c|c|c|c|c|c|} 
\hline
       \text{Figure} & $N$ &$\| d \|$  & $\lambda$ \\
 \hline
       \ref{fig:ERPLnet}(a-b)   &   $10^4$ & $100$, $50$, $30$, $20$, $10$  & $101$, $51.1$, $31.1$, $20.1$, $11.1$ \\ \hline
     \ref{fig:ERstdP}(a-b), \ref{fig:R_vs_K_theoretical}(a) & $10^4$ &$100$  &  $101$  \\ \hline
      \ref{fig:ERerror}(a) & Varied & Varied  & - \\ \hline
      \ref{fig:ERerror}(b) & $10^4$ & Varied & - \\ \hline
      \ref{fig:growthRate} & $2.5 \times 10^4$ & $5000$ & $4999$\\ \hline
\end{tabular}} 
   \caption{Parameters for the Erd\"os-Renyi (ER) networks studied here. Here $N$ is the number of nodes, $\| d \|$ is the mean degree, and $\lambda$ is the largest eigenvalue of $A$. In all cases, the correlation coefficient, \Eq{rhodef}, is  $\rho = 1$.
}
\label{tab:booktabs}
}
\end{table*}

\begin{table*}[htb]
\centering{
\scalebox{0.95}{
   \begin{tabular}{|c|c|c|c|c|c|c|c|}
\hline
       \text{Figure} & $\alpha$  & $d_{\min}$  & $\lambda$ & $\rho$ \\
 \hline

     \ref{fig:ERPLnet}(c-d),  \ref{fig:maxSynchrony}   & $2.5$, $2.8$,  $3.1$, & $33$, $44$,  $52$,  & $299$, $243$, $201$, & $0.52$, $0.81$,  $0.84$, \\ 
     & $3.5$, $3.8$  & $60$, $64$ & $162$, $140$  & $0.93$, $0.99$ \\ \hline 
     \ref{fig:PLcorr}(a-b) & 2.5 & 33.3 & $282$, $364$, $459$ & $0.78$, $0.97$, $1.09$ \\ \hline
       \ref{fig:R_vs_K_theoretical}(b)  &  $2.8$  & $44$ & $243$ & 0.81 \\ \hline
\end{tabular}} 
   \caption{Parameters for the scale-free (SF) networks studied here. Here $\alpha$ and $d_{min}$ are parameters in \Eq{powerlaw}, $\lambda$ is the largest eigenvalue of $A$, and $\rho$ is the correlation coefficient \Eq{rhodef}. In all cases the number of oscillators is $N = 10^4$ and the mean degree is $\| d \| = 100$. 
}
\label{tab:booktabs2}
}
\end{table*}

With $R_n=0$, \Eq{dMomentum_dT_orderParam} implies that $\dot{p}_n = 0$ or $p_n = \bar{p}_n =$ constant and each oscillator rotates with a constant angular frequency. Let us denote these solutions by
\begin{align*} 
	\bar{\theta}_n(t) &= \bar p_n t + \bar\theta_n^0 \enspace,\\
	\bar{p}_n(t) &= \bar p_n \enspace.
\end{align*}
We will assume that the initial phases $\bar\theta_n^0$ are uniformly distributed in $[0,2\pi)$.
Letting $(\delta \theta_n, \delta p_n)$ denote small perturbations to the incoherent state 
($\bar\theta_n(t), \bar p_n(t))$, linearizing \Eq{dPhase_dT}-\Eq{dMomentum_dT} gives
\begin{equation}
\begin{split}\label{eq:linearizedODEs}
	\delta \dot{\theta}_n &= \delta p_n \enspace, \\
	\delta \dot{p}_n &= \frac{K}{N} \sum_{m=1}^{N} A_{nm} \cos\left(\bar\theta_m-\bar\theta_n\right) \delta \theta_m \enspace,
\end{split}
\end{equation}
upon neglecting $(K/N)\sum_{m} A_{nm} \cos\left(\bar\theta_m- \bar\theta_n \right) = \mathcal{O}(\sqrt{d}/N)$.
These equations can be solved for the fastest growing mode using the new variables
\[
	B_n(t) = \sum_{m=1}^N A_{nm}e^{i\theta_m(t)}
		[\delta \theta_n(t)-\delta\theta_n(t_0)] \enspace.
\]
As we show in App.~\ref{dispersion}, upon setting $B_n(t) = b_n e^{st}$, with the complex growth rate $s = \gamma + i\omega$, and assuming that $\gamma > 0$, then in the limit $t \to \infty$ the eigenvector $\{b_n\}$ and growth rate are determined by the eigenvalue problem 
\begin{align} \label{eq:dispersion}
	b_k = \frac{K}{2N} \sum_{n=1}^{N} \frac{A_{kn}b_n}{(s-i\bar{p}_n)^2} \enspace.
\end{align}
Equivalently, $2N/K$ is an eigenvalue of the matrix $A \,\diag\{(s-i\bar{p}_n)^{-2}\}$. For a given matrix $A$, distribution of initial momenta $\bar{p}_n$, and coupling constant $K$, \Eq{dispersion} determines the growth rate $\gamma$ and oscillation frequency $\omega$ of perturbations from the incoherent state.

\begin{figure*}[t]
\centering{ \subfloat{{\label{fig:ER_K}}\includegraphics[width=0.4\textwidth]{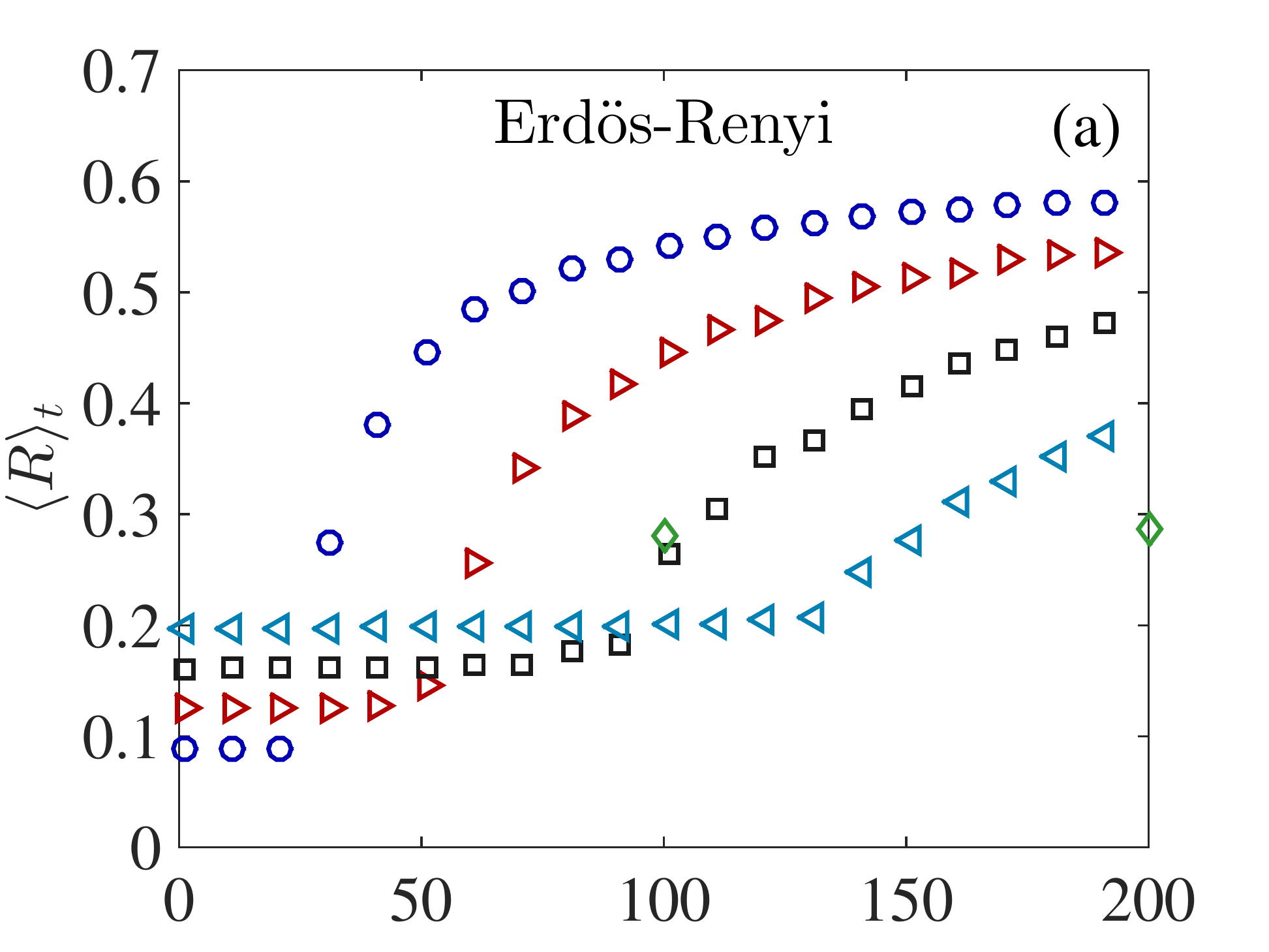}}
\subfloat{{\label{fig:ER_K_lambda}}\includegraphics[width=0.4\textwidth]{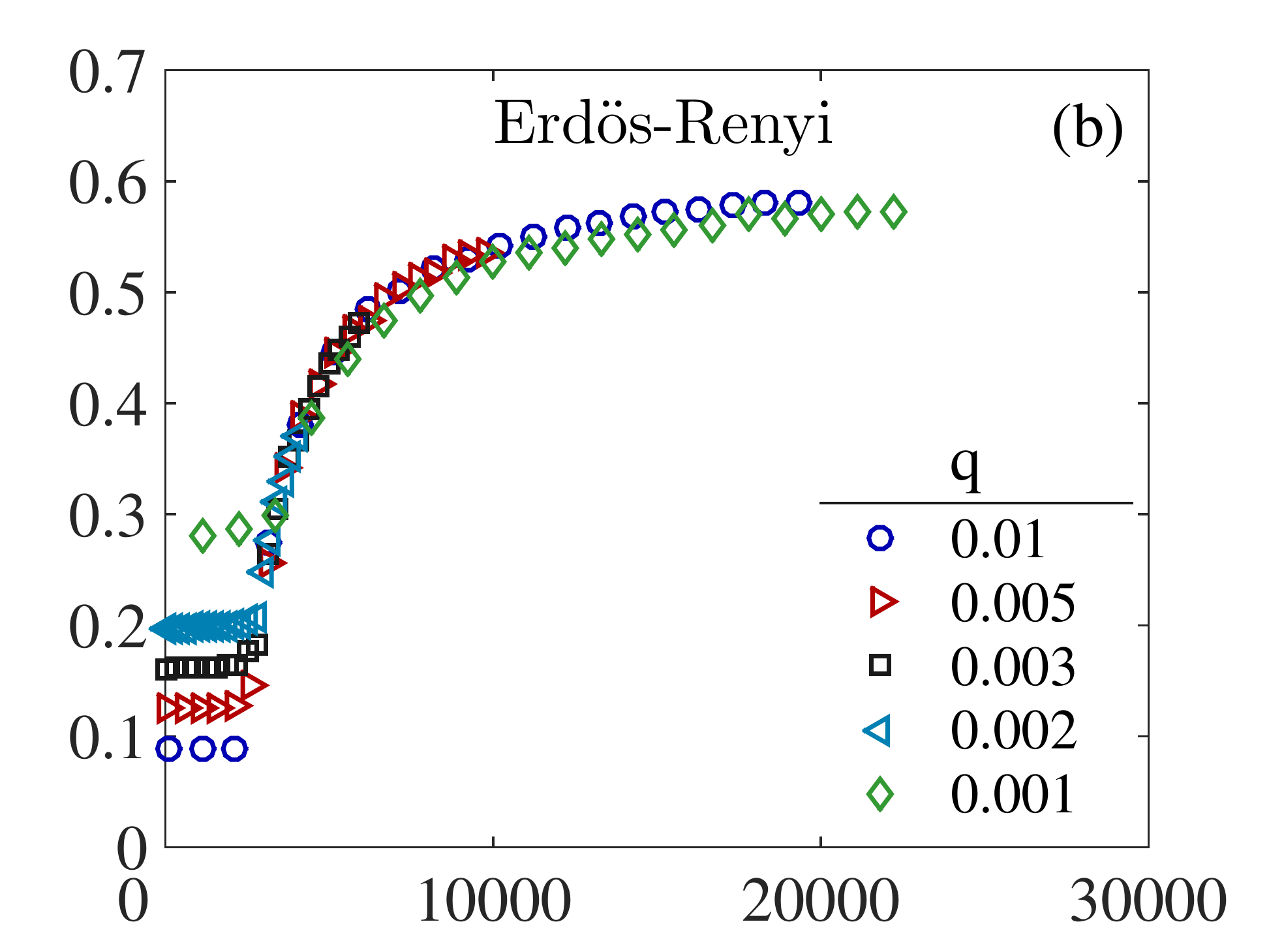}}\\
\subfloat{{\label{fig:PL_K}}\includegraphics[width=0.4\textwidth]{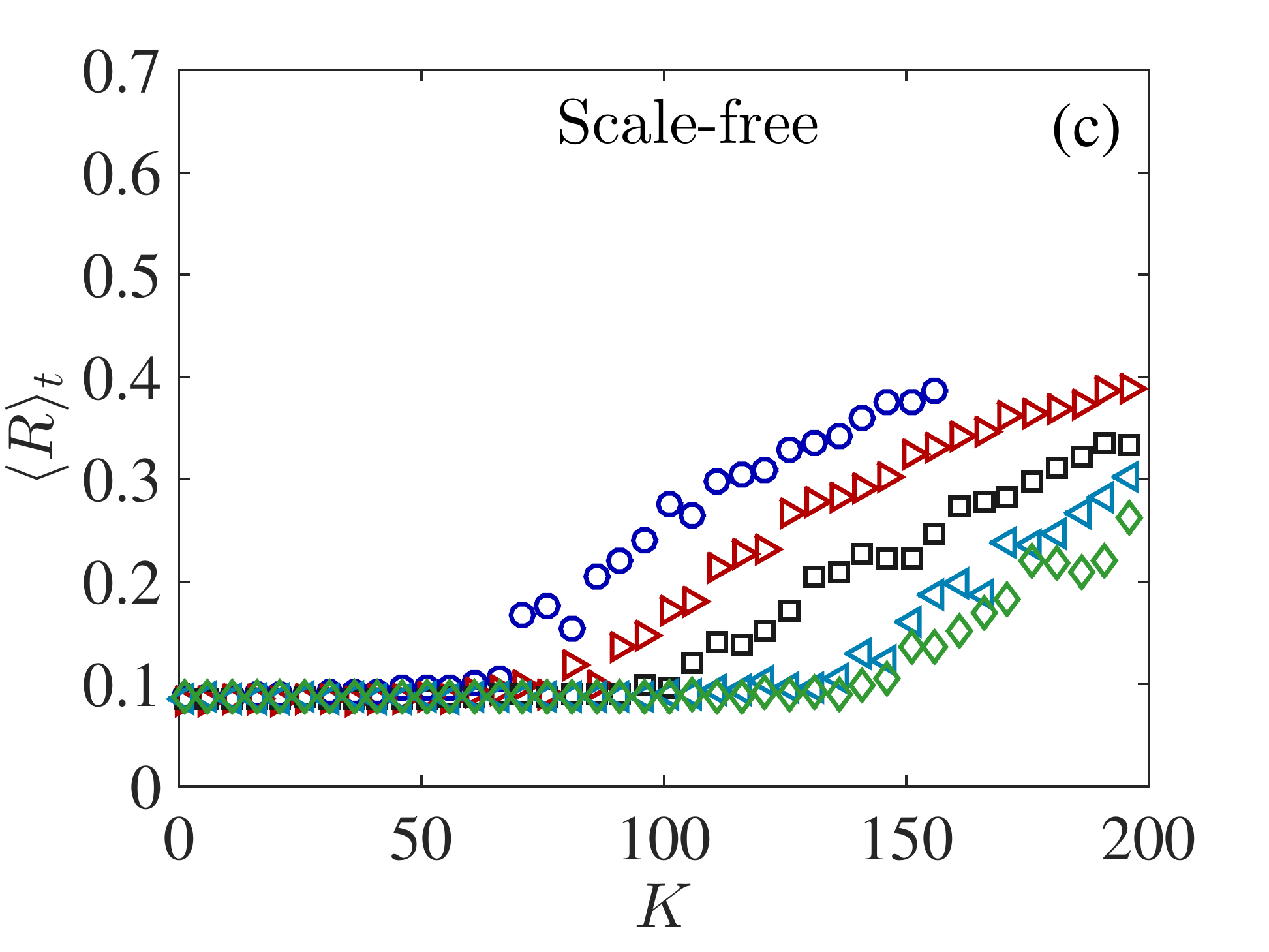}}
\subfloat{{\label{fig:PL_K_lambda}}\includegraphics[width=0.4\textwidth]{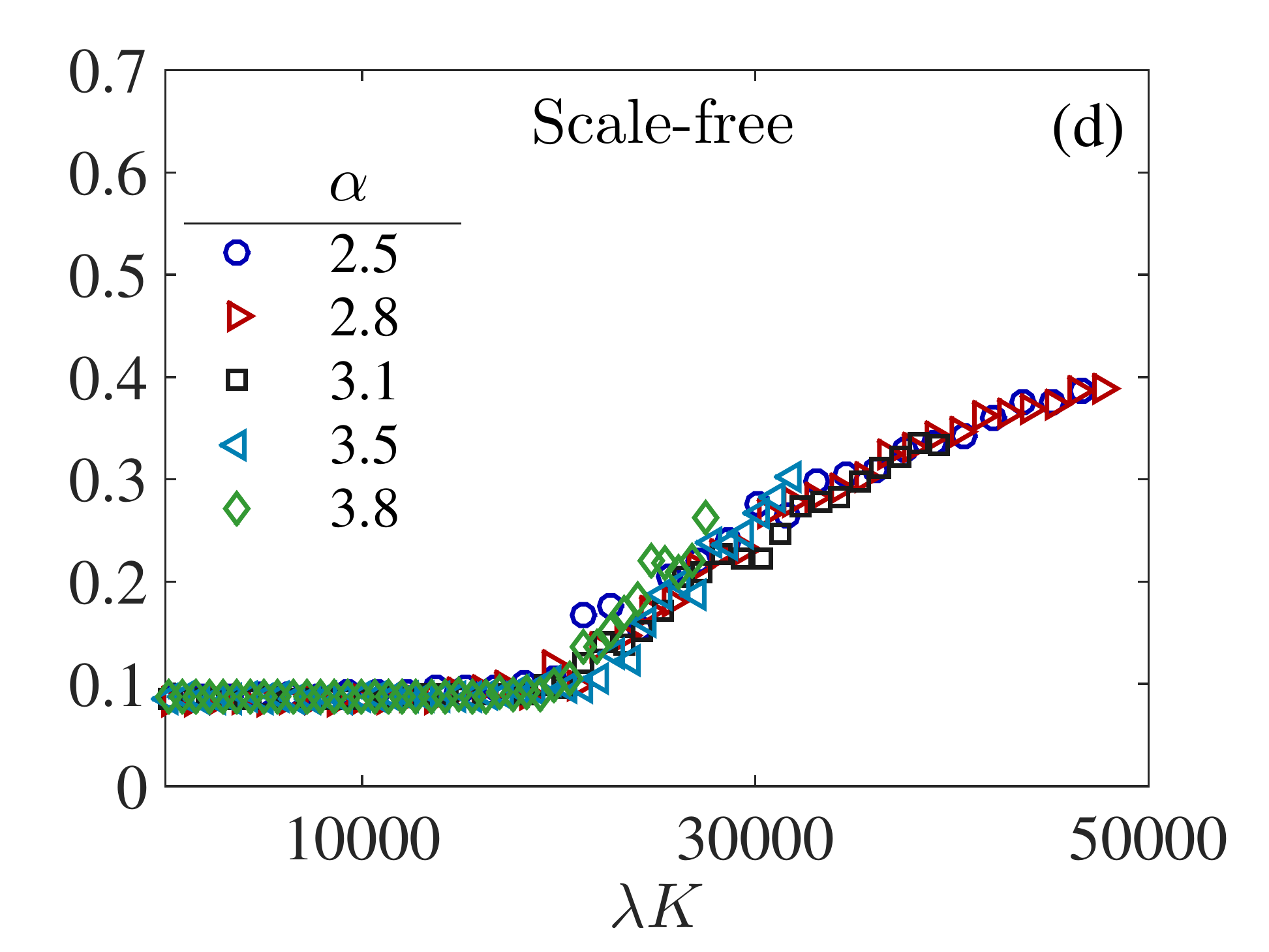}} \\
}
\caption{Time average of the order parameter $R$, i.e., $\langle R \rangle_t$
as a function of $K$ [(a) and (c)] and $\lambda K$ [(b) and (d)] for a variety of network structures having $N=10^4$. Panels (a) and (b) show the results for simulated Erd\"os-Renyi networks with varying $q$. Panels (c) and (d) show the results for simulated scale-free networks with varying $\alpha$. From $(b)$ and $(d)$, we see that plotting $\langle R \rangle_t$ against $\lambda K$ for any network causes the transitions to line up, in agreement with \Eq{criticalKFirstOrder}.}
\label{fig:ERPLnet}
\end{figure*}

In the rest of the paper, we will consider---for simplicity---the case in which the initial momenta $\bar{p}_n$ are independent of the network properties, i.e., of $A$. That is, we can consider the set $\{(A_{kn}, \bar{p}_n): n=1,\ldots, N\}$ to be a sample from a joint distribution that is, in fact,  a product of two independent distributions, one for the network ($A$) and one for the initial conditions ($\bar{p}$). In this case, we propose to look for solutions $\{b_n\}$ of \Eq{dispersion} that are also statistically independent of the momenta. This hypothesis will be verified, a posteriori, below. Since the mean of a product of functions of independent variables is the product of their means, we can approximate \Eq{dispersion} by
\begin{equation} \label{eq:bkApprox}
	b_k 
		\approx \frac{K}{2} \left\| \frac{1}{(s-i\bar{p})^2} 
				\right\| \frac{1}{N}\sum_{n=1}^{N} A_{kn} b_n \enspace.
\end{equation}
This is an eigenvalue equation; indeed, suppose that $\lambda$ is an eigenvalue of $A$ and $\{b_n\}$ its corresponding eigenvector, then \Eq{bkApprox} gives
\begin{align} \label{eq:lambdaHat}
	1 = \frac{K \lambda}{2N} \left\| \frac{1}{(s-i\bar{p})^2} \right\| \;.
\end{align}
This verifies the hypothesis: if the momenta are uncorrelated with $A$, they will also be uncorrelated with its eigenvectors $\{b_n\}$, thus justifying our derivation of  \Eq{bkApprox}.
The eigenvector that corresponds to the earliest onset of instability (i.e., the smallest $K$) is that corresponding to the eigenvalue of $A$ with largest magnitude, which we will henceforth denote just by $\lambda$ (we assume $A$ is nonnegative, irreducible and aperiodic so that $\lambda$ is unique by the Perron-Frobenius theorem).  In what follows we study the growth rate associated with this mode. If the momenta $\bar p$ have the distribution $g(\bar{p})$ we can write, in the limit $N \to \infty$,
\begin{align*} 
	\frac{2N}{K \lambda} = \int_{-\infty}^{\infty} \frac{g(p) \mathrm{d}p}{(s - ip)^2} \enspace.
\end{align*}
Integrating by parts,
\begin{align*} 
	\frac{2N}{K \lambda} = i \int_{-\infty}^{\infty} \frac{g'(p) \mathrm{d}p}{s - ip} \enspace,
\end{align*}
where $g' = dg/dp$. Now let $s = \gamma + i\omega$. Inserting this and separating real and imaginary parts and noting
that $\lambda$ is real since $A$ is symmetric, we get:
\begin{equation} \label{eq:gDashPSGammaPlusOmega}
\begin{split}
	\frac{2N}{K \lambda} &= \int_{-\infty}^{\infty} \frac{g'(p) (\omega - p) \mathrm{d}p}
		{\gamma^2 + (\omega - p)^2} \enspace, \\
	0 &= \int_{-\infty}^{\infty} \frac{g'(p) \mathrm{d}p}{\gamma^2 + (\omega - p)^2} \enspace.
\end{split}
\end{equation} 
As an example, we consider the case in which $g(p)$ is a Gaussian centered at $p = \Omega$ with standard deviation $\sigma_0$. By symmetry, the second equation is satisfied when $\omega = \Omega$. The first equation in \Eq{gDashPSGammaPlusOmega} then yields
\begin{align} \label{eq:gammaKKc}
	\frac{\gamma}{\sigma} \sqrt{\frac{\pi}{2}} e^{\frac{\gamma^2}{2\sigma_0^2}} 
		\mathrm{erfc}\left(\frac{\gamma}{\sqrt{2}\sigma}\right)  
		 = 1 - \frac{2 \sigma_0^2 N} {\lambda K} \enspace.
\end{align}

We find the critical coupling strength $K_c$ by letting $\gamma \to 0^+$, obtaining
\begin{align} \label{eq:criticalKFirstOrder}
	K_c &= \frac{2 \sigma_0^2 N} {\lambda} \enspace.
\end{align}
The dependence of $K_c$ on the largest eigenvalue of $A$ is similar to that observed in various other dynamical systems on networks such as the Kuramoto model \cite{restrepo:ott:hunt:2005}, epidemic spreading \cite{mieghem:2012}, and the propagation of avalanches \cite{larremore:et:al:2012}. The largest eigenvalue captures various effects of network structure including the degree distribution and degree-degree correlations \cite{restrepo:ott:hunt:2007}.

To get the growth rate $\gamma$ for any given $K > K_c$, we can invert equation \Eq{gammaKKc} numerically. We note that, given a value of $K/K_c$,  the growth rate $\gamma$ is independent of the structure of network.

\subsection{Numerical experiments}
\label{subsec:numericalExperiments}

\begin{figure*}[t]
\centering{
\subfloat{\includegraphics[width=0.4\textwidth]{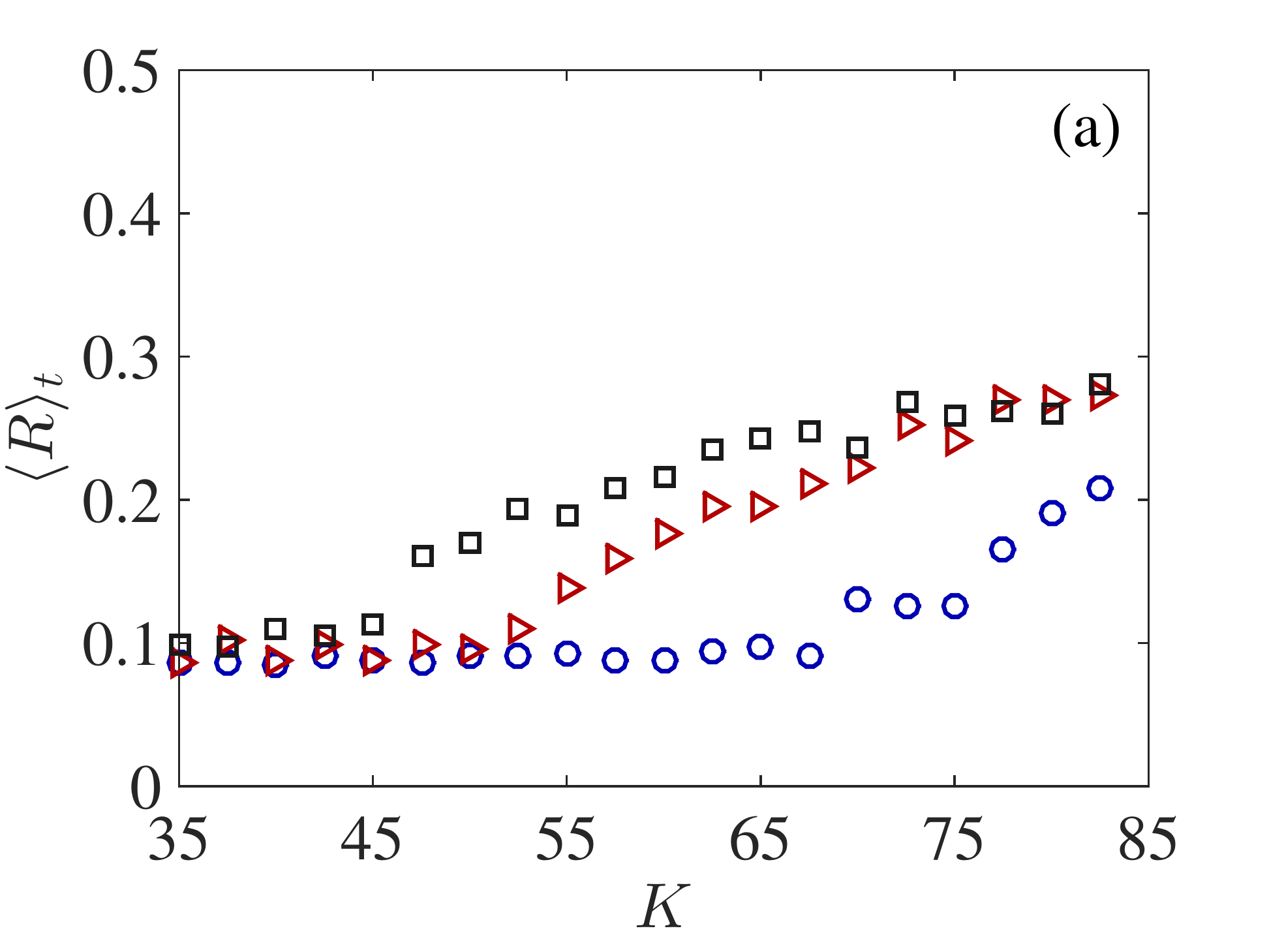}}
\subfloat{\includegraphics[width=0.4\textwidth]{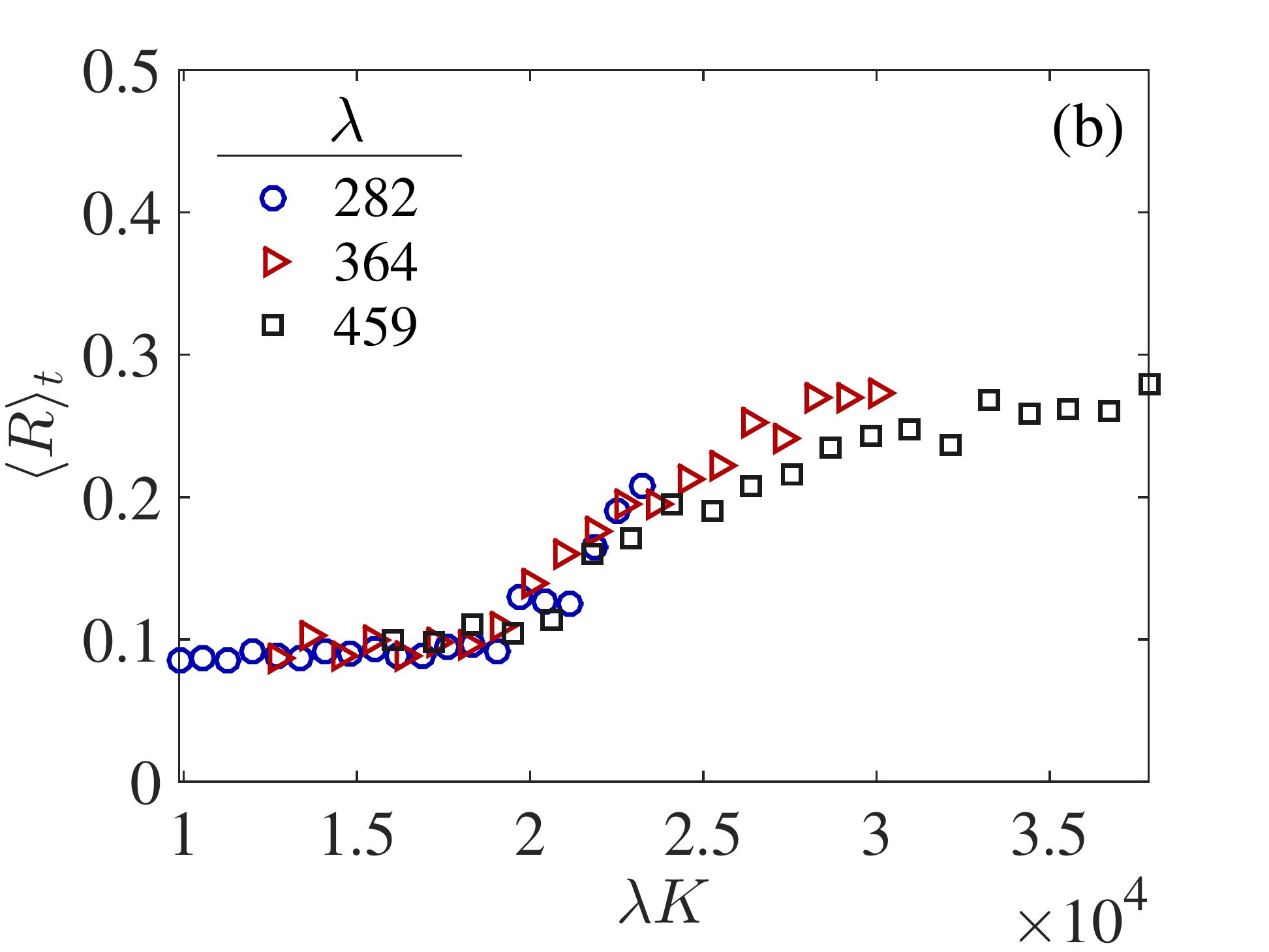}}
}
\caption{Time averaged global order parameter $R$ as a function of $K$ and $\lambda K$ for scale-free networks with varying edge-degree correlation $\rho$. The three networks have distinct $\lambda$ [the inset of panel (b)], and panel (a) shows they have distinct $K_c$. Panel (b) plots $\langle R \rangle_t$ against $\lambda K$ showing that the three curves nearly coincide as they did in Fig.~\ref{fig:ERPLnet}.}
\label{fig:PLcorr}
\end{figure*}

\begin{figure*}[htb]
\centering{
\subfloat{{\label{fig:ER_K_vary_stdP}}\includegraphics[width=0.4\textwidth]{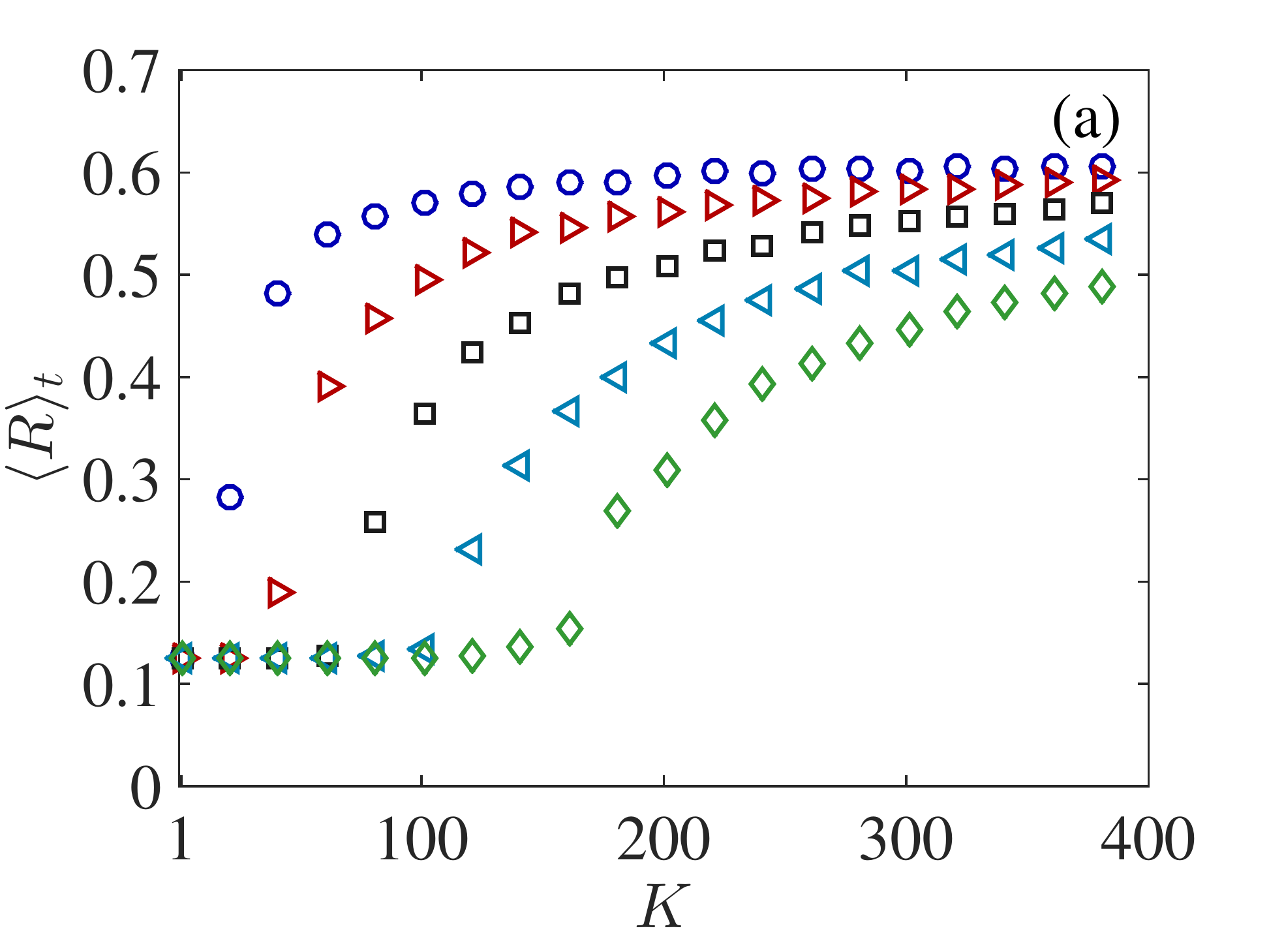}}
\subfloat{{\label{fig:ER_K_by_stdP}}\includegraphics[width=0.4\textwidth]{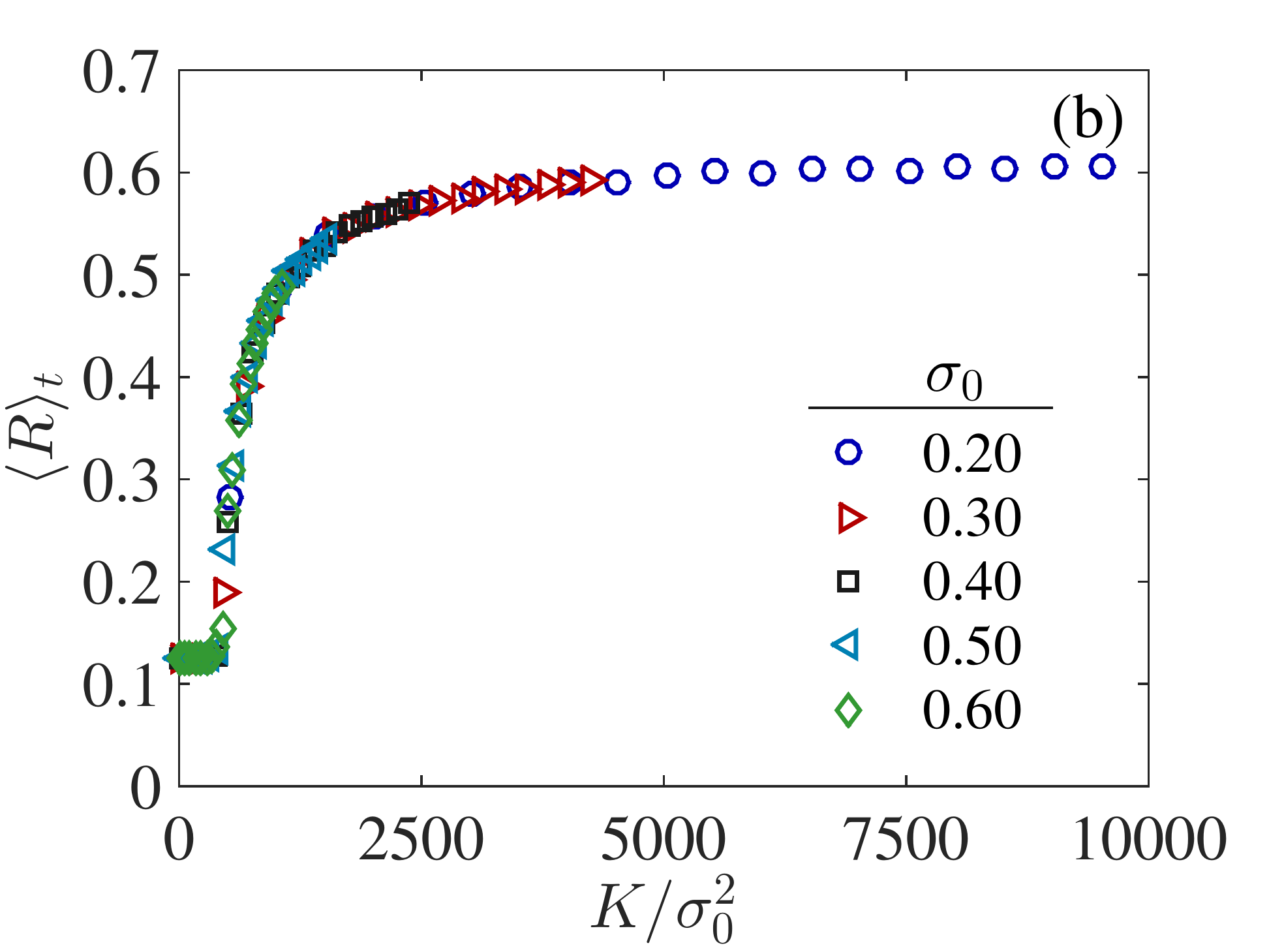}}
}
\caption{Time-averaged global order parameter for a fixed Erd\"os-Renyi network as a function of (a) $K$ and (b) $K / \sigma_0^2$. The initial momentum distribution is a Gaussian with mean $\Omega = 6$ and varying standard deviation $\sigma_0$, as shown in the inset of (b).}
\label{fig:ERstdP}
\end{figure*}

Here we present computations for several examples to validate and illustrate our results and verify some of the assumptions in the derivations of \Sec{dispersion} for the linear stability of the incoherent state. As network examples, we use both Erd\"os-Renyi and scale-free networks with $N$ nodes. We generate the Erd\"os-Renyi networks by establishing an undirected link between nodes $n$ and $m$ (i.e., setting $A_{nm} = 1$) with probability $q$, and not establishing a link ($A_{nm} = 0$) with probability $1-q$ to obtain a network with mean degree $\|d \| = (N-1)q$.
The scale-free networks are generated, using the algorithm of Chung and Lu \cite{chung:lu:vu:2003}, to have a target degree distribution of the form
\begin{equation}\label{eq:powerlaw}
	P(d) = \left\{ \begin{array}{cc}
		 d^{-\alpha}  & d > d_{\min} \enspace, \\
		  0           & d \le d_{min} \enspace.
		 	\end{array} \right.
\end{equation}
Given a value of $\alpha$, we choose $d_{\min}$ to achieve a desired mean degree $\| d \|$, which will be noted for each specific case. Tables~\ref{tab:booktabs} and \ref{tab:booktabs2} show the parameters used in the various experiments for the two types of networks.

In the following set of experiments, we fix $N=10^4$ for both network types. We start with the phases $\theta_n(0)$ uniformly distributed in $ [0, 2\pi)$. The initial momenta $p_n(0)$ are sampled from a Gaussian distribution with mean $\Omega=6$ and a standard deviation $\sigma_0=0.35$. We integrate Eqs. \eqref{eq:dPhase_dT} and \eqref{eq:dMomentum_dT_orderParam} using a second-order \emph{leap-frog} algorithm with time step $h = 0.01$. In most of our experiments we report a time averaged value of the global order parameter, i.e., $\langle R \rangle_t$ as a function of the coupling strength, $K$. Integrations start at $K = 0$, and $K$ is periodically incremented by $\Delta K$ (in the plots, $\Delta K$ is the separation between consecutive symbols). Integration at the new $K$ value continues from the current state. The total integration time for each value of $K$ is typically $T_2 = 1000$ time units; this includes a period during which transients decay, typically $T_1 =500$ time units. The time average of $R$ is computed using \Eq{timeAverage}.

\begin{figure*}[htb]
\centering{
\subfloat{{\label{fig:ER_r_vs_N}}\includegraphics[width=0.4\textwidth]{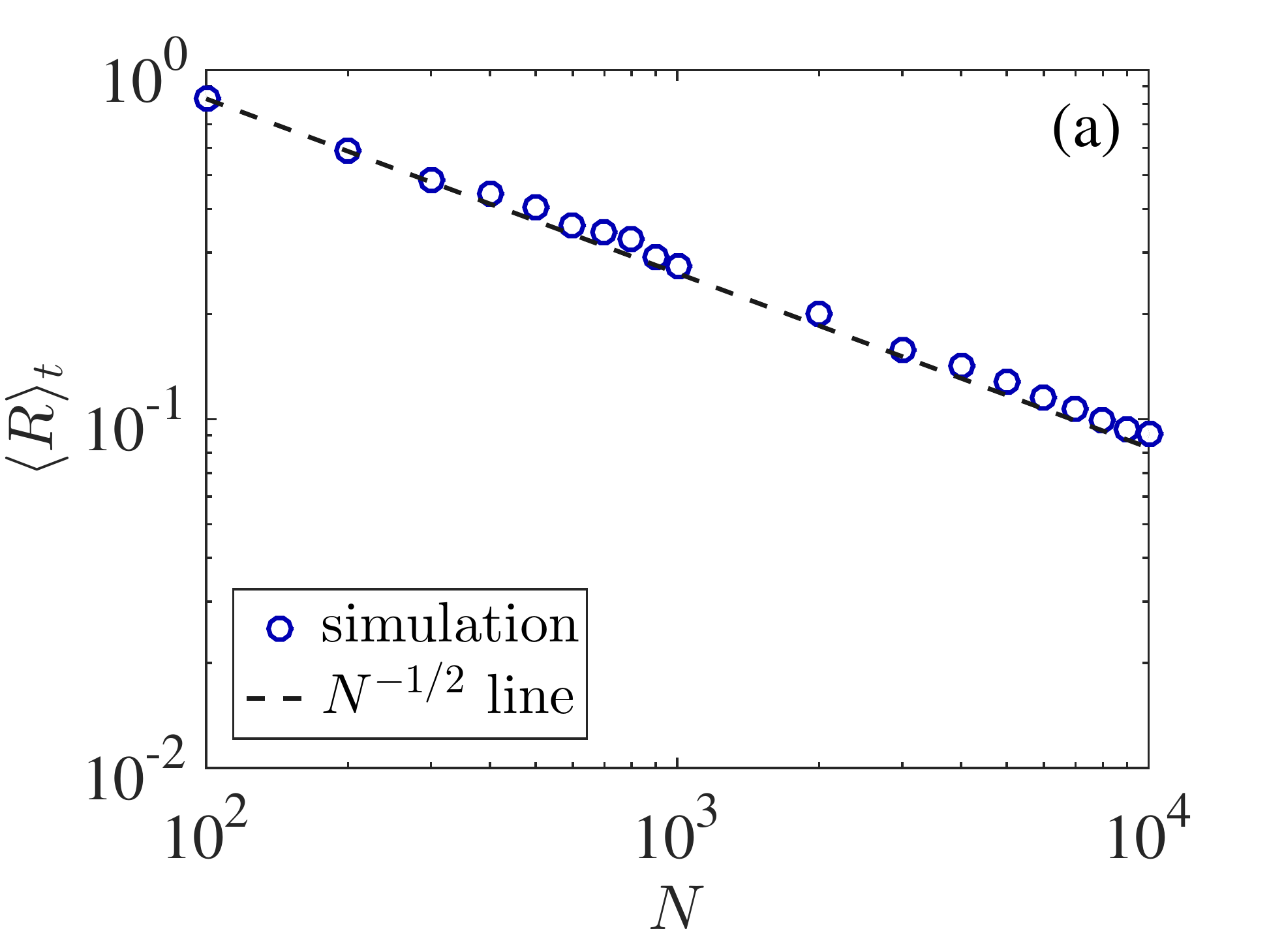}}
\subfloat{{\label{fig:ER_r_vs_p}}\includegraphics[width=0.4\textwidth]{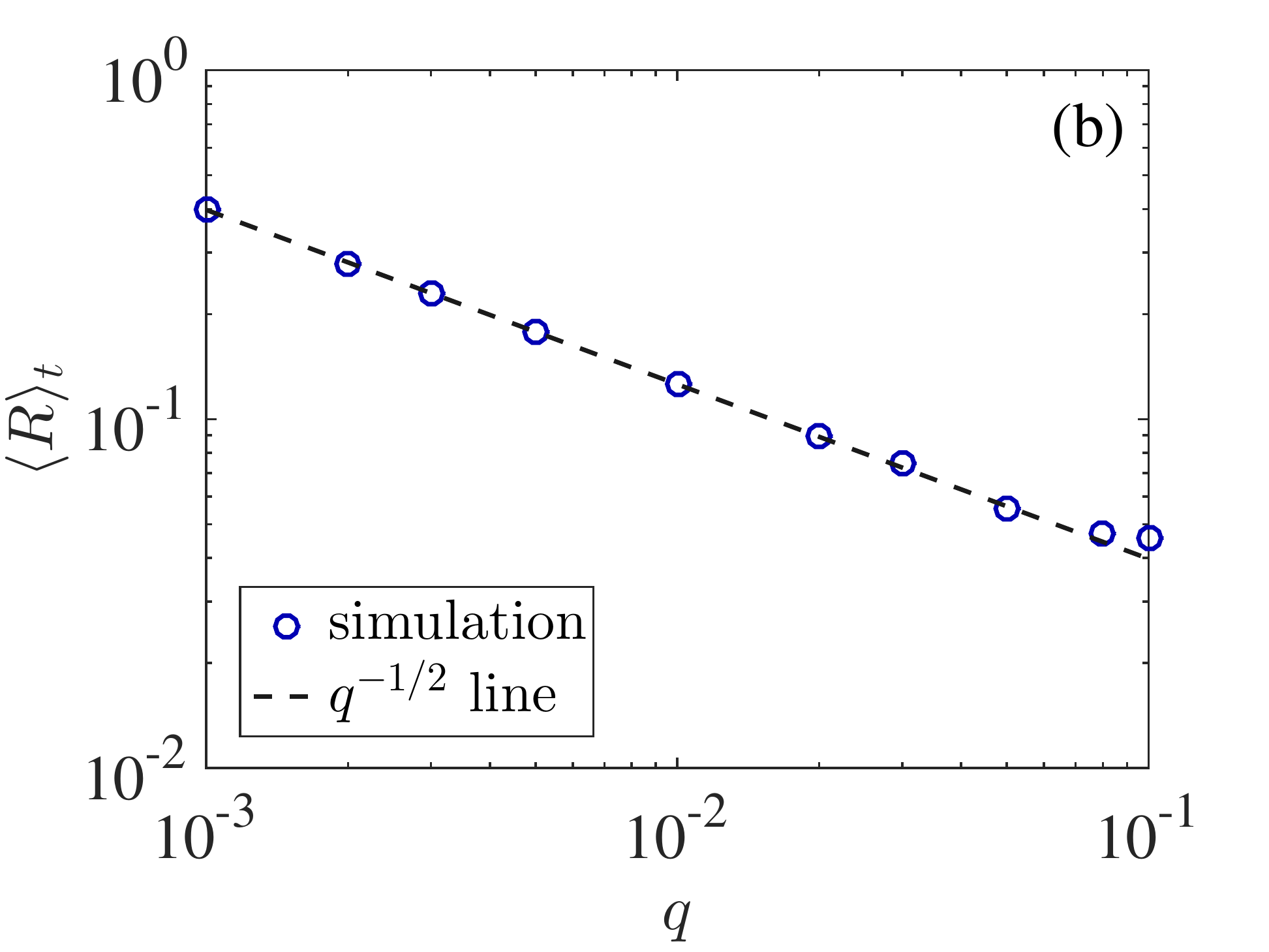}}
}
\caption{Time averaged global order parameter as a function of (a) network size, $N$, and (b) edge probability, $q$ for Erd\"os-Renyi networks with $K = \tfrac12 K_c$. The dashed lines have a slope of $-\tfrac12$ and an arbitrary intercept, corresponding to the theoretical estimate $\langle R \rangle_t \sim (qN)^{-1/2}$.}
\label{fig:ERerror}
\end{figure*}

The first experiment studies the effect of varying the link probability $q$  for the Erd\"os-Renyi networks and of varying the degree exponent $\alpha$ \Eq{powerlaw} for the scale-free networks; results are shown in Fig.~\ref{fig:ERPLnet}. Panel (a) shows the time averaged global order parameter \Eq{globalOrderParam} as a function of the coupling strength $K$ for Erd\"os-Renyi networks with various values of $q$ [indicated in the inset of Fig.~\ref{fig:ERPLnet}(b)], and (c) shows the same quantity for scale-free networks with various values of $\alpha$ [indicated in the inset of Fig.~\ref{fig:ERPLnet}(d)] with $\| d \| = 100$. These networks have different largest eigenvalues, $\lambda$ (recall Tbls.~\ref{tab:booktabs}-\ref{tab:booktabs2}), and therefore, in agreement with \Eq{criticalKFirstOrder}, synchronization begins at different values of $K$. However, for all of the networks the onset of the transition to synchrony begins at the same value of $\lambda K$, see panels $(b)$ and $(d)$, confirming that $K_c \propto 1/\lambda$.
Remarkably, the entire set of curves collapse onto a single curve, indicating that even the partially synchronized states depend on the network structure only through $\lambda$.

As a second experiment, we use scale-free networks to study the effect of increasing degree-degree correlations. Correlations between the degrees $d_n$, $d_m$ 
of the nodes connected by a randomly chosen edge, $A_{nm} \neq 0$, (also known as assortative mixing by degree \cite{newman:2003}) can modify the eigenvalue $\lambda$ which, by \Eq{criticalKFirstOrder}, should affect the onset of synchrony. These correlations can be quantified by the coefficient  \cite{restrepo:ott:hunt:2007}
\begin{equation}\label{eq:rhodef}
	\rho = \frac{\| d_n d_m \|_e}{\| d_n\|_e \| d_m\|_e} \enspace,
\end{equation}
where $\| \cdot \|_e$ denotes an average over the edges.  In our experiments we first construct a scale-free network with $d_{\min} = 33$ and $\alpha=2.5$; then we rewire the edges following the algorithm of \cite{restrepo:ott:hunt:2007} to increase $\rho$. The initial network has $\rho = 0.78$; subsequent rewiring creates an intermediate network with $\rho = 0.94$, and a final network with $\rho = 1.09$. These three networks also have different principal eigenvalues, recall Tbl.~\ref{tab:booktabs2}; however, all have the same degree distribution. The results, in Fig.~\ref{fig:PLcorr} (a), show  the onset of synchronization occurs at different values of $K$, as in the first experiment. Therefore the simple scaling $K_c \propto 1/\| d\|$ is not sufficient to describe the behavior of heterogeneous networks like scale-free networks. However,  when $\langle R \rangle_t$ is plotted against $\lambda K$, as shown in Fig.~\ref{fig:PLcorr} (b), the transition points again align as predicted by \Eq{criticalKFirstOrder}.

In a third numerical experiment we study the effect of varying the distribution of initial momenta $p_n$. More specifically, using a single Erd\"os-Renyi network with $N=10^4$ and $q=0.01$, we consider a Gaussian distribution of momenta $g(p)$ with mean $\Omega = 6$ and various standard deviations, $\sigma_0$. Figure~\ref{fig:ERstdP}(a) shows a plot of $\langle R\rangle_t$ versus $K$ for the different values of $\sigma_0$ and panel (b) shows the collapsed version when the abscissa is $K /\sigma^2$. As expected from \Eq{criticalKFirstOrder}, the critical values collapse to one point in the latter case, and---as before---the entire set of curves nearly coincide near the transition.

As mentioned before, in the incoherent state we expect that $R_n$ has fluctuations of order $\sqrt{d_n}/N$. Therefore $R = \sum_{n=1}^{N} R_n / \|d\|$ should scale as $R \sim \| \sqrt{d}\| / \|d\|$. For Erd\"os-Renyi networks the distribution is sharply peaked about $\| d \| \approx N q$. Therefore we should have that $R \sim (qN)^{-1/2}$. To verify that the observed finite value of $R$ is consistent with these finite-size effects, we varied the network size $N$ and edge probability $q$, holding the coupling constant fixed at $K = \tfrac12 K_c$. The results are shown in Fig.~\ref{fig:ERerror}. The time average $\langle R \rangle_t$ is shown as a function of $N$ for $q=0.01$ in panel (a), and as function of $q$  for $N = 10^4$ in panel (b). These results show that $\langle R \rangle_t \sim N^{-1/2}$ and $\langle R \rangle_t \sim q^{-1/2}$, respectively (the dashed lines have a slope of $-\tfrac12$ with arbitrary intercept), consistent with fluctuations due to the finite size of the network.

\begin{figure}[htb]
\centering{
\includegraphics[width=0.4\textwidth]{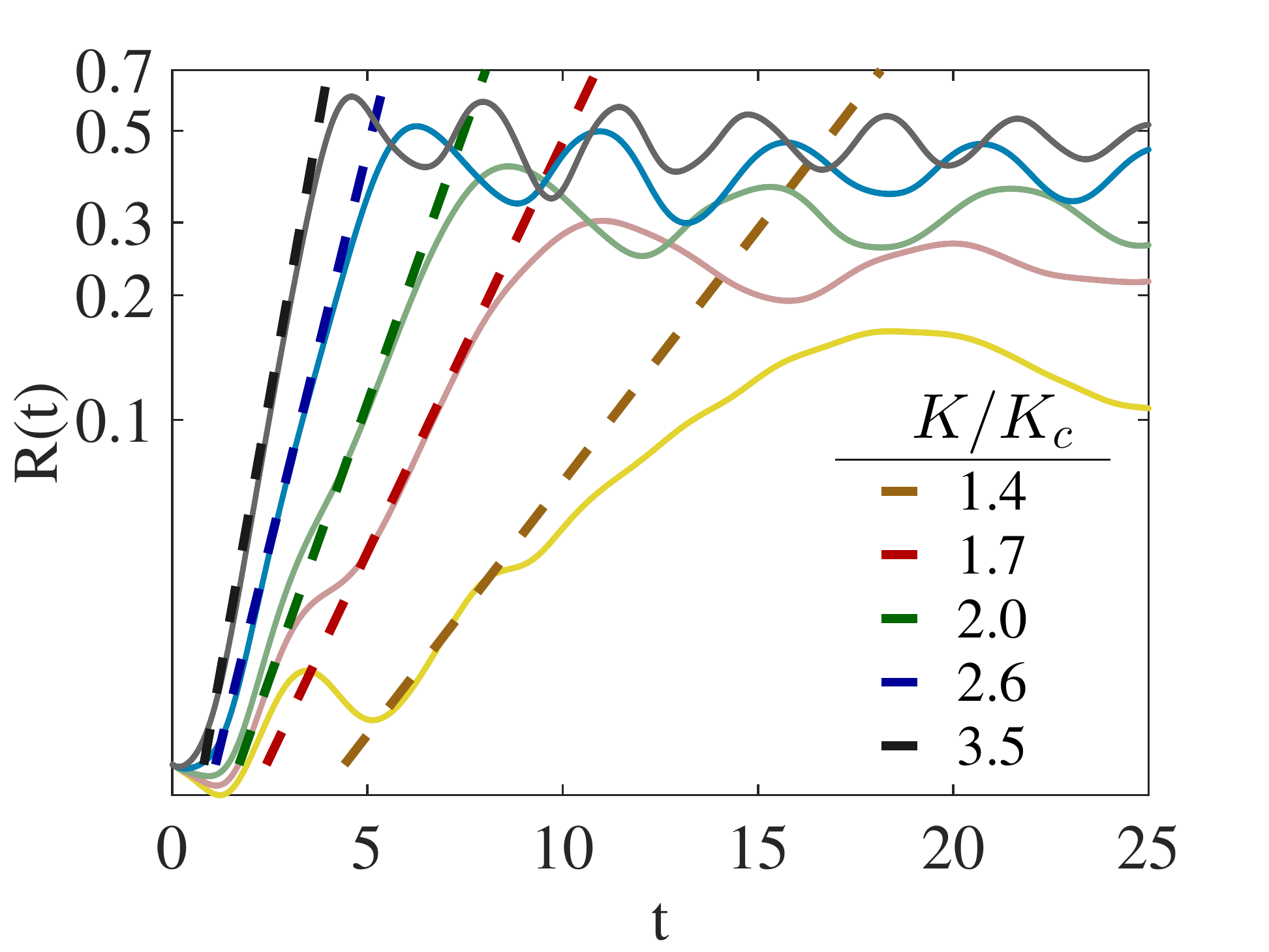}
}
\caption{We study the growth of the order parameter $R$ on a log scale over time plotted on the $x$-axis. The solid lines show simulated data for an Erd\"os-Renyi network and the dashed lines show the corresponding theoretical result from \Eq{gammaKKc}. The different curves are for a variety of $K$ values. Theory and simulations agree well with increasing $K/K_c$ values.}
\label{fig:growthRate}
\end{figure}

Finally, we test our results for the linear growth rate $\gamma$ from \Eq{gammaKKc}. Here we use  an Erd\"os-Renyi network with $N=25,000$ and $\| d \| = 5000$ (the reason for the larger $N$ and $\|d\|$  is discussed below). We plot $R(t)$ on a log scale as a function of $t$ in Fig.~\ref{fig:growthRate}. The solid lines represent data from direct numerical integration of \Eq{dPhase_dT}-\Eq{dMomentum_dT}, and the dashed lines have the slope $\gamma$ predicted from \Eq{gammaKKc}. 

Of course, exponential growth predicted by the linear theory can occur only when $R \ll 1$.
In addition, \Eq{gammaKKc} gives the growth rate of the fastest growing mode, but initial conditions may contain a mixture of different modes. Therefore, we expect the theoretical growth rate only over an intermediate time domain where the fastest growing mode dominates, but where $R(t)$ has not yet saturated. To make his region as large as possible within our computational constraints we chose $\| d \| = 5000$. Since a quantitative comparison would require to arbitrarily select an interval [$R_{\mathrm{MIN}}, R_{\mathrm{MAX}}$] to compute a slope, we present here just the curves in Fig.~\ref{fig:growthRate} and do not attempt to fit portions of these curves. Despite these difficulties, the simulations of the full model show a growth rate that seems to be well approximated by \Eq{gammaKKc}.

\section{Synchronized state}
\label{sec:synchronizedState}

\begin{figure*}[htb]
\centering{
\subfloat{{\label{fig:ER_K_vs_r_theoretical}}\includegraphics[width=0.4\textwidth]{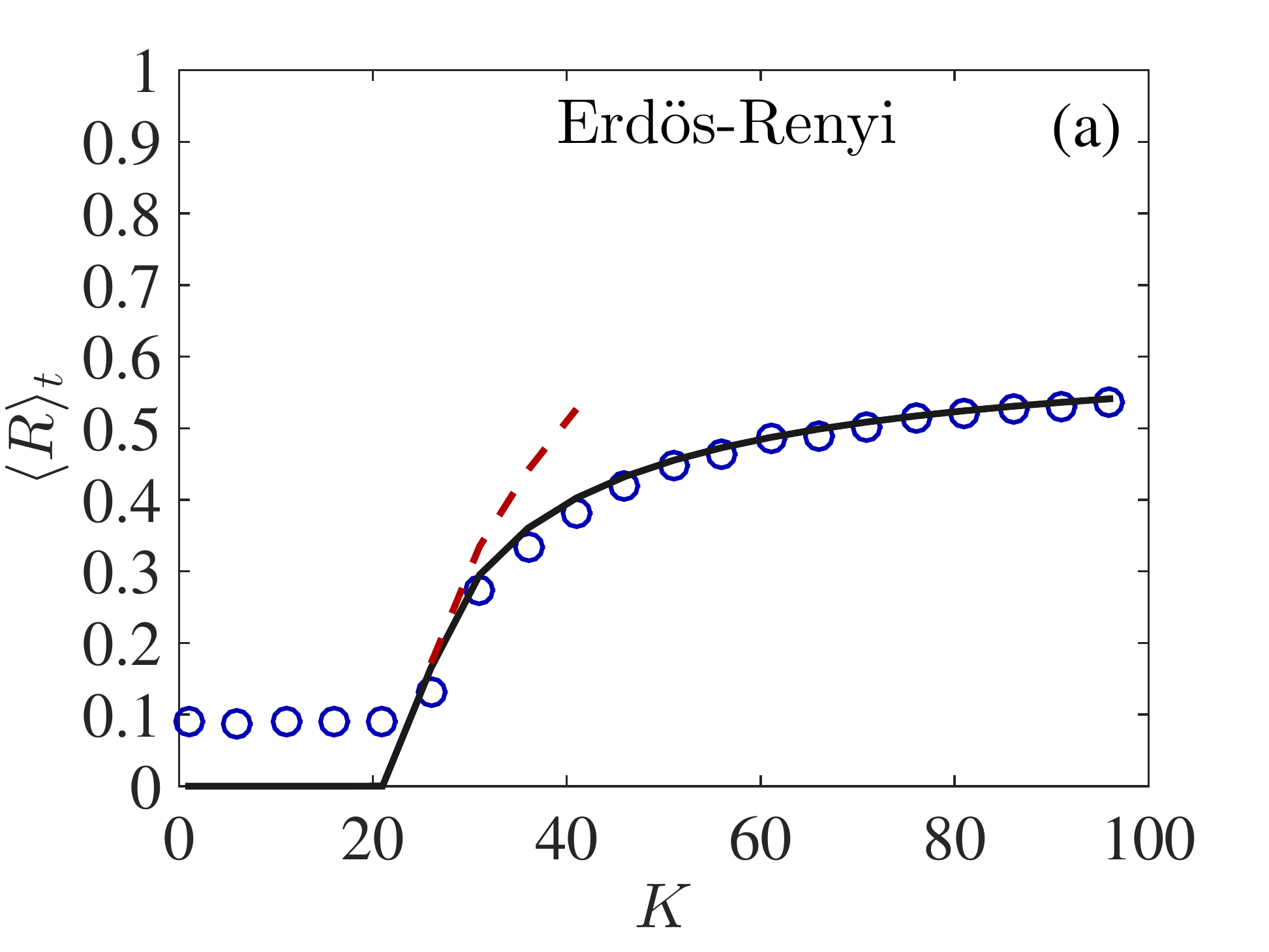}}
\subfloat{{\label{fig:PL_K_vs_r_theoretical}}\includegraphics[width=0.4\textwidth]{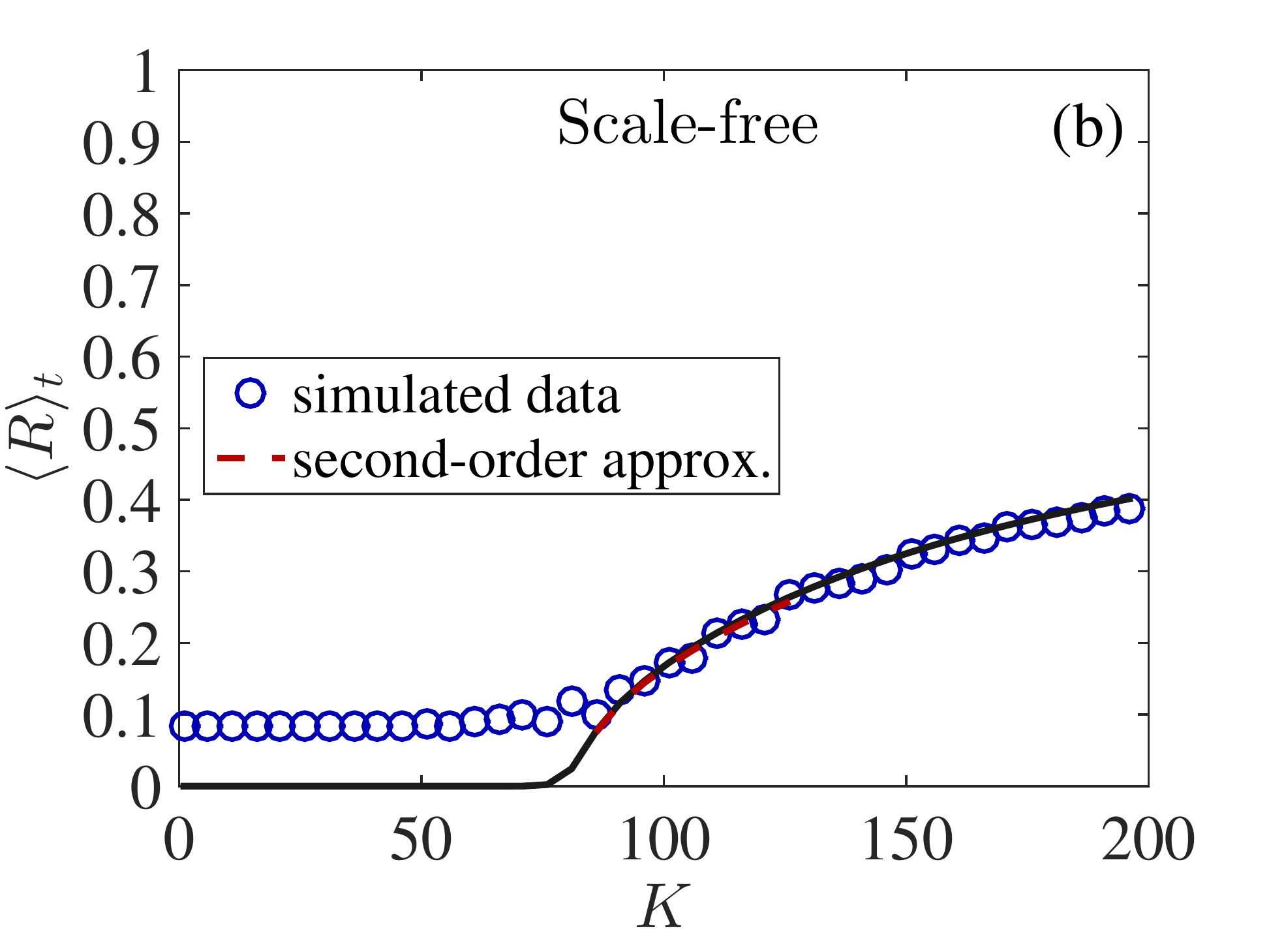}}
}
\caption{Time averaged global order parameter, $\langle R \rangle_t$ as a function of the coupling strength $K$  for (a) Erd\"os-Renyi ($q=0.01$) and (b) scale-free ($\alpha = 2.5$, $d_{min} = 33$) networks with $N=10^4$ and $\| d \|= 100$. 
The circles (blue) are the result of numerical integration for an initial Gaussian distribution $g_0(p)$ with mean $P_0 = 6$ and variance $\sigma_0^2 = 0.12$ (for Erd\"os-Renyi) and $\sigma_0^2 = 1$ (for scale-free) and random phases uniformly distributed in $[0,2\pi)$. The solid curves (black) are the numerical solutions to \Eq{rSigmaBesselMain}. The dashed curves (red) show the second-order approximation \Eq{globalOrderRInDeltaK}, valid near $K=K_c$ (see \Sec{approxOnsetSynchrony}).
}
\label{fig:R_vs_K_theoretical}
\end{figure*}

In \Sec{linearStabilityAnalysis}, we studied  the incoherent state; now we turn our attention to synchronized solutions, i.e., solutions for which the local order parameters $R_n$ are nonzero even in the $N\to \infty$ limit. We are interested in the long-time average of the order parameters \Eq{localOrderParam}, 
$
  \langle R_n \rangle_t
$. 
Following \cite{restrepo:meiss:2014} and based on numerical experiments (see below), we look for solutions such that the different local order parameters exhibit, on average, a phase synchrony---they rotate with a common frequency $\Omega$ and at a common phase $\vphi$:
\[
	R_n e^{i \psi_n} \approx r_n e^{i \left(\Omega t + \vphi\right)} \enspace,
\]
where $r_n$ is constant. This is a nontrivial assumption that we expect to be valid when all the nodes have neighbors that are representative of the network as a whole and the distribution of momenta is sufficiently narrow. 
To implement this assumption, we write
\begin{equation} \label{eq:RnTermsOfrnAndzn}
	R_n e^{i\psi_n} = (r_n + z_n)e^{i(\Omega t + \vphi)} \enspace,
\end{equation}
where the real average local order parameters $r_n$ and global phase $\vphi$ are defined by
\begin{equation}\label{eq:rnDef}
	r_n e^{i\vphi} \equiv \langle R_n e^{i\left(\psi_n - \Omega t\right)}\rangle_t.
\end{equation}
so that $\langle z_n \rangle_t = 0$.
There are two implicit assumptions here: first, there is an $\Omega$ such that $|\langle z_n(t) \rangle_t| \ll r_n$, so that $z_n$ represents the fluctuations,
and second, there is a single phase $\vphi$ that makes all of the $r_n$ real. The goal of this section is to obtain equations that will determine both the common frequency $\Omega$ and the local order parameters $r_n$.

As in \cite{restrepo:meiss:2014}, we define new variables $\bar\theta_n$ and $\bar{p}_n$ in a rotating frame,
\begin{equation} \label{eq:thetaPBar}
\begin{split}
	\bar\theta_n &= \theta_n - (\Omega t + \vphi) \enspace,\\
	\bar{p}_n &= p_n - \Omega \enspace .
\end{split}
\end{equation}
Inserting these and \Eq{RnTermsOfrnAndzn} into \Eq{dPhase_dT} and \Eq{dMomentum_dT_orderParam} gives
\begin{align} 
	\dot{\bar\theta}_n &= \bar{p}_n \enspace,  \label{eq:changeThetaNBar}\\
	\dot{\bar{p}}_n &= -K r_n \sin(\bar\theta_n) + K \Im (z_n e^{-i \bar\theta_n}) \enspace.  \label{eq:changePNBar}
\end{align}
The second term of \Eq{changePNBar} can be thought of as a perturbation to the Hamiltonian dynamics of each oscillator that preserves the total energy of all the oscillators \Eq{hamiltonianNetworkCase}. We treat this perturbation as if it were stochastic and assume that the probability of observing node $n$ in a given region of the phase space $\left(\theta_n, p_n\right)$ over a long time is given by a Boltzmann distribution \cite{gibbs:1902}. More precisely, we assume that for any function $f$ of the single oscillator variables, for large starting time $T_1$ and large interval $T_2-T_1$, the time average \Eq{timeAverage} limits to a phase space average:
\begin{align} \label{eq:boltzmannAssumption}
	\langle f \rangle_t \to \langle f \rangle_g \equiv 
	 \int_{0}^{2\pi} {\int_{-\infty}^{\infty} f \left(\bar\theta_n, \bar{p}_n \right) 
		g\left(\bar\theta_n, \bar{p}_n; r_n\right) \, 
			\mathrm{d}\bar{p}_n} \, \mathrm{d}\bar\theta_n \enspace.
\end{align}
Here $g$ is the Boltzmann distribution for the single-rotor energy 
\begin{align} \label{eq:gFunction}
	g(\bar\theta, \bar{p}; r) = \frac{\beta^{1/2}}{(2\pi)^{3/2} \mathrm{I}_0(K \beta r) } e^{-\beta \left(\bar{p}^2/2 - Kr \cos (\bar{\theta})\right)} \enspace,
\end{align}
for an inverse temperature $\beta$ that must be determined. The Bessel function, $\mathrm{I}_0$, in the denominator normalizes
the distribution:
$\int_{0}^{2\pi} \int_{-\infty}^{\infty} g\left(\bar\theta, \bar{p}; r\right)\, \mathrm{d}\bar{p} \, \mathrm{d}\bar{\theta} = 1$. For this distribution, the mean square momentum (in this case, the variance of $\bar{p}$) is
\begin{equation} \label{eq:betaInTermsSigmaSq}
	\sigma^2 =  \int_{-\infty}^{\infty}\int_{0}^{2\pi} \bar{p}^2\, g(\bar{p}, 
	\bar{\theta};r) \, \mathrm{d}\bar{\theta}\, \mathrm{d}\bar{p}  = \beta^{-1} \enspace,
\end{equation}
and the mean potential energy is proportional to 
\begin{equation}\label{eq:cosThetaBar}
	\int_{-\infty}^{\infty} \int_{0}^{2\pi} \cos(\bar{\theta}) \, g(\bar{\theta}, \bar{p}; r)  \, \mathrm{d}\bar{\theta}\, \mathrm{d}\bar{p}
	= v(K \beta r) \enspace,
\end{equation}
where we introduce the notation
\begin{equation}\label{eq:BesselRatio}
	v(x) \equiv \frac{\mathrm{I}_1(x)}{\mathrm{I}_0(x)} \enspace,
\end{equation}
and $\mathrm{I}_1$ is the first order Bessel function.

Using \Eq{rnDef} and \Eq{thetaPBar} in the definition \Eq{localOrderParam} of the local order parameter, we can solve for $r_n$, and then use \Eq{boltzmannAssumption} and
\Eq{cosThetaBar} to obtain
\begin{equation}\label{eq:rnBessel}
	r_n = \frac{1}{N} \sum_{m=1}^{N} A_{nm} \langle \cos(\bar\theta_m) \rangle_t 
	  = \frac{1}{N} \sum_{m=1}^{N} A_{nm} v(K \beta r) \enspace.
\end{equation}

Equation \Eq{rnBessel} depends on the inverse temperature $\beta$ introduced in \Eq{gFunction}, which can be determined by conservation of energy. 
Suppose that initially the rotors have a distribution of momenta with mean $P_0 = \|p_n(0)\|$ and variance $\sigma_0^2 = \|(p_n(0) - P_0)^2\|$, and that they have a distribution of phases $\theta_n(0)$ with potential energy $U_0 = -K/(2N) \sum_{n,m} A_{nm} \cos(\theta_m(0) - \theta_n(0))$.
The initial energy  is then
\begin{align} \label{eq:Hsub0}
	E_{0}   = \frac{N}{2} \left( P_0^2 + \sigma_0^2 \right) + U_0 \enspace .
\end{align}
Since the total momentum is conserved by \Eq{dMomentum_dT}, the mean momentum at any time remains equal to $P_0$.
Under the Boltzmann assumption \Eq{boltzmannAssumption}, the mean $\langle \bar{p} \rangle_g = 0$, which, by \Eq{thetaPBar}, implies that
\begin{equation}\label{eq:MomentumConserved}
	P_0 = \langle \Omega + \bar{p} \rangle_g = \Omega \enspace .
\end{equation}

In the new coordinates \Eq{thetaPBar}, the total energy \Eq{hamiltonianNetworkCase} at any time is
\[ 
	E =  \frac{1}{2} \sum_{n=1}^{N} {\left(\Omega + \bar{p}_n\right)}^2 - 
     \frac{K}{2} \sum_{n=1}^{N}  r_n  \cos\left(\bar\theta_n\right) 
     - \frac{K}{2} \sum_{n=1}^{N}  \Re(z_n e^{-i\bar \theta_n})\enspace ,
\]
Since the energy is constant, we can take a time average and use \Eq{betaInTermsSigmaSq} to obtain
\begin{equation}\label{eq:timeAveragedHV2}
	E = \frac{N}{2} (\Omega^2 + \sigma^2) -  \frac{K}{2} \sum_{n=1}^{N}  r_n  \langle \cos\left(\bar\theta_n\right) \rangle_t - \frac{K}{2} \sum_{n=1}^{N}  \Re \langle z_n e^{-i\bar \theta_n} \rangle_t
\end{equation}
We now neglect the terms proportional to the fluctuations $z_n$ (see below for a discussion). Since energy and momentum are conserved, $E_0 = E$ and $P_0 = \Omega$, we can apply the Boltzmann assumption \Eq{boltzmannAssumption} and combine \Eq{cosThetaBar}, \Eq{Hsub0}, and \Eq{timeAveragedHV2} to compute the variance:
\begin{align} \label{eq:sigmaSqEq}
	\sigma^2 = \sigma_0^2 + \frac{K}{N} \sum_{n=1}^{N} r_n v(K \beta r) + \frac{2}{N}U_0 
		\enspace .
\end{align}
Substituting for $\beta$ using \Eq{betaInTermsSigmaSq} in \Eq{rnBessel} and \Eq{sigmaSqEq} gives a closed system of $N+1$ equations for the local order parameters and the variance: 
\begin{equation} \label{eq:rSigmaBesselMain}
\begin{split}
	r_n &= \frac{1}{N} \sum_{m=1}^{N} A_{nm} v\left(\frac{K r_m}{\mu}\right) \enspace, \\
	\sigma^2& = \sigma_0^2 + \frac{K}{N} \sum_{n=1}^{N} r_n v\left(\frac{K r_m}{\mu}\right) + \frac{2}{N}U_0  \enspace.  
\end{split}
\end{equation}

This system generalizes analogous self-consistent results for the all-to-all coupled case (e.g., see Eq.~(16) in \cite{antoni:ruffo:1995}) Note that this system always has the trivial, incoherent solution $r_n = 0$, $n= 1,\ldots, N$ and $\sigma^2 = \sigma_0^2 + 2U_0/N$. By the analysis of \Sec{dispersion}, this solution is stable when $K < K_c$. We note that when the initial conditions are in the incoherent state, i.e., when the phases are uniformly distributed in $[0,2\pi)$, the potential energy term $2U_0/N$ is negligible in the limit $N \to \infty$.

\begin{figure*}[htb]
\centering{
\subfloat{{\label{fig:OmegaDistribution}}\includegraphics[width=0.4\textwidth]{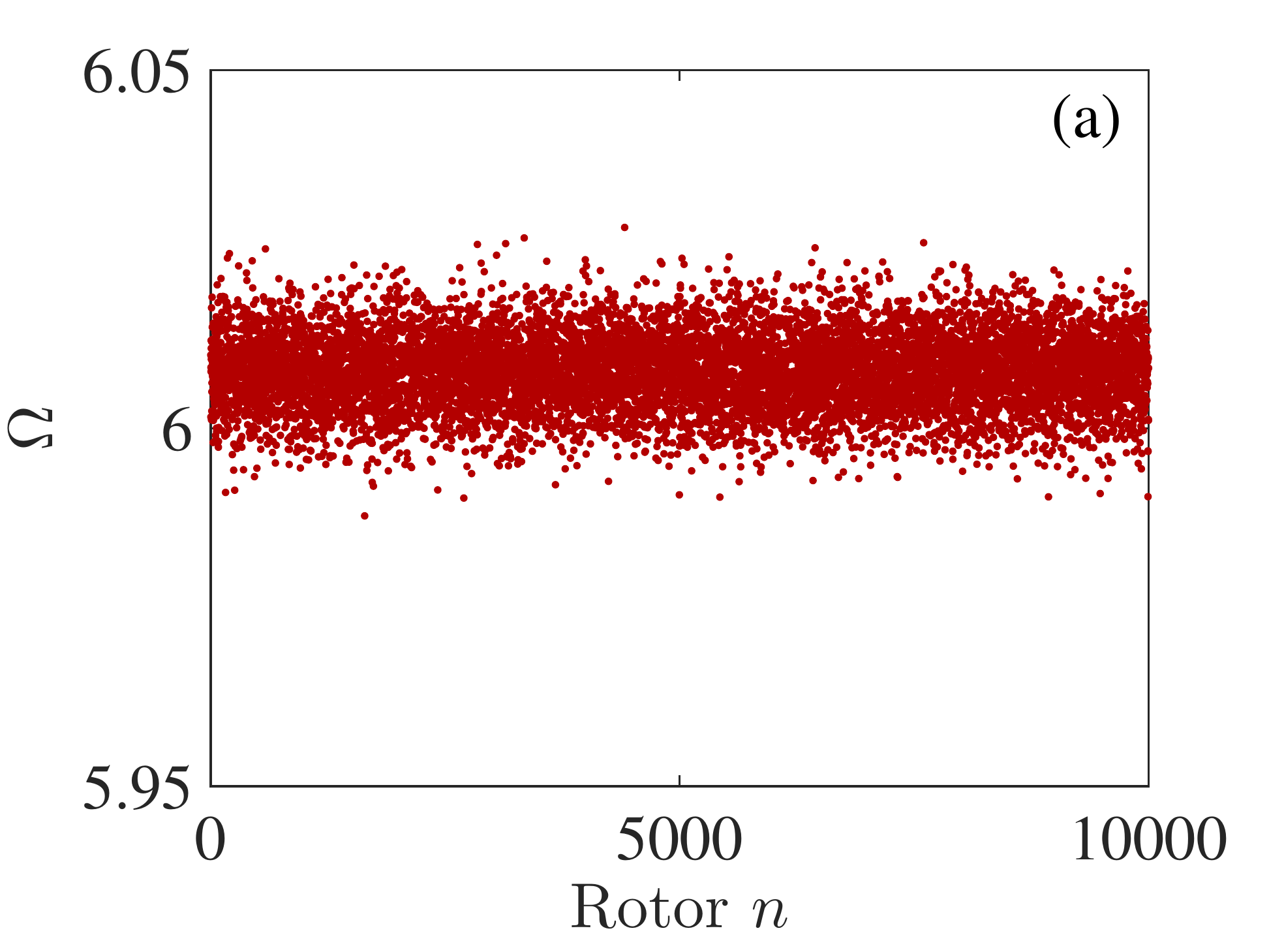}}
\subfloat{\label{fig:znIndependentOfThetaBarNAssumption}\includegraphics[width=0.4\textwidth]{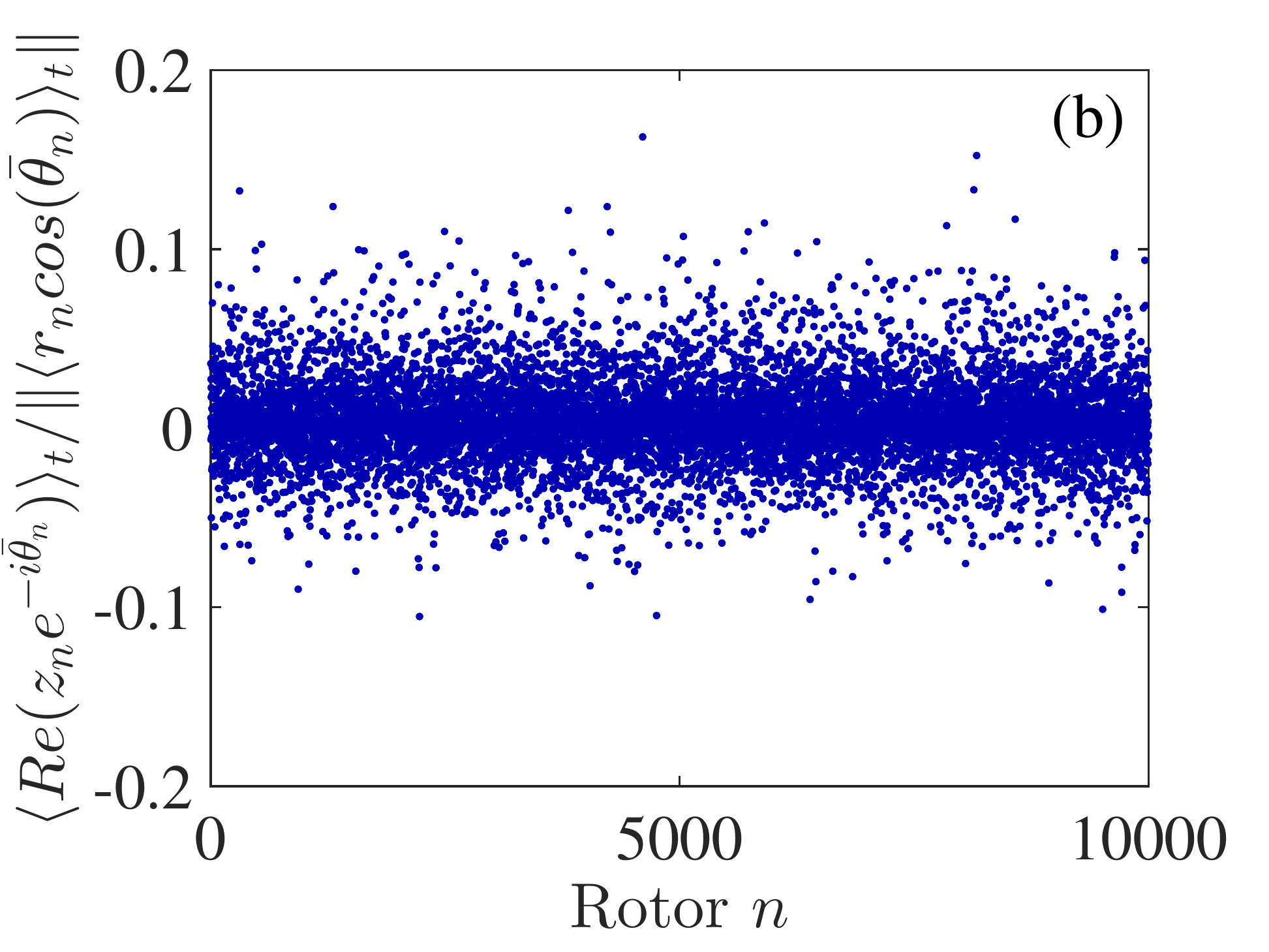}}
}
\caption{Time averaged oscillator frequencies, panel (a), and fluctuation amplitudes, panel (b), for oscillators in an Erd\"os-Renyi network ($q=0.01$, $\| d \| = 100$, $N=10^4$), with $K = 600 = 3K_c$. The initial momenta had a Gaussian distribution with mean $P_0 = 6$, and variance $\sigma_0^2 = 1$. Time averages were taken over an interval $[2000,10000]$. Panel (b) shows the time average of the fluctuating terms in \Eq{timeAveragedHV2} relative to the synchronized terms.}
\label{fig:Assumptions1}
\end{figure*}

\begin{figure}[b]
\centering{
\includegraphics[width=0.4\textwidth]{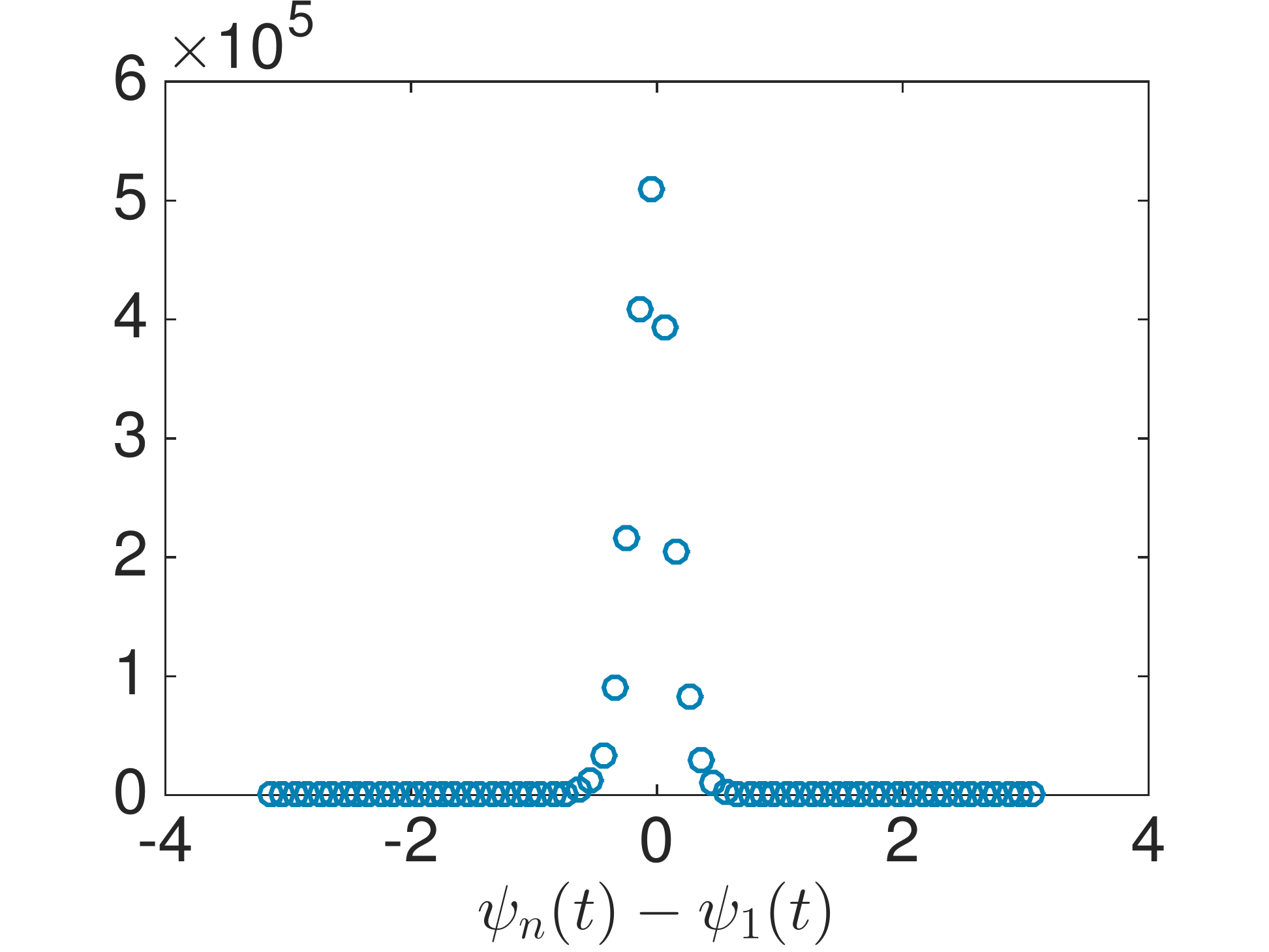}
}
\caption{Histogram of $\psi_n(t) - \psi_1(t)$, for the network of Fig.~\ref{fig:OmegaDistribution}, but with $N = 10^3$ oscillators. Here the $\psi$ are sampled each unit of time over a total of 2000 time units, so the total number of events is $2(10)^6$.}
\label{fig:histPsinMinusPsi1}
\end{figure}

To find a nontrivial, synchronized solution with $r_n >0$, we solve the system \Eq{rSigmaBesselMain} numerically for the $N+1$ variables $\{r_n\}$ and $\sigma^2$. A simple method is fixed point iteration: given a guess $\{r_n >0\}$ and $\sigma^2$, new values can be computed from \Eq{rSigmaBesselMain} using the guesses on the right-hand sides. Numerically, this iteration converges to values that appear to be independent of the initial guess, suggesting that there is a unique solution to these equations, and that there is a nontrivial solution, $r_n > 0$, when $K > K_{c}$.
Once the local order parameters are known, the global order parameter is computed from \Eq{globalOrderParam}.

In Fig.~\ref{fig:R_vs_K_theoretical} we show a comparison of the predictions of 
\Eq{rSigmaBesselMain} (black solid lines) with direct numerical integration of the ODEs \Eq{dPhase_dT}-\Eq{dMomentum_dT} (blue circles) for both an Erd\"os-Renyi network, panel (a), and a scale-free network, panel (b). 
The theory agrees with the simulations, except that when $K < K_c$ the observed order parameter $R$ is not zero, as predicted by the theory, because of finite-size effects.

Before moving on to the next section, we present a discussion of the three main assumptions made to derive \Eq{rSigmaBesselMain}:
\begin{assumption}{1}{}\label{assum:1}
The local order parameters have a common rotation frequency and phase, $\psi_n = \Omega t + \vphi$ [introduced in \Eq{RnTermsOfrnAndzn}].
\end{assumption}
\begin{assumption}{2}{}\label{assum:2}
The final state is ergodic and has a Boltzmann distribution \Eq{boltzmannAssumption}
[used in \Eq{rnBessel} and \Eq{timeAveragedHV2}].
\end{assumption}
\begin{assumption}{3}{}\label{assum:3}
Fluctuations can be neglected: $\| \langle \Re(z_n e^{-i\bar \theta_n}) \rangle_t\|  \ll \| \langle r_n  \cos\left(\bar\theta_n\right) \rangle_t \|$ [for \Eq{timeAveragedHV2}].
\end{assumption}

The first assumption, that the phases of the local order parameters are all the same, is reasonable when the network is constructed in such a way that the neighbors of different nodes have the same statistical properties and the initial momentum distribution is sufficiently narrow (since $\dot{\theta}_n = p_n$).
Indeed, this assumption has also been used successfully in studies of Kuramoto oscillators on complex networks \cite{arenas:et:al:2008, restrepo:ott:hunt:2005}. It is expected to break down for networks where the oscillator properties are correlated with the network structure, such as lattices with spatially dependent frequencies \cite{acebron:et:al:2005} and communities with different oscillator properties \cite{arenas:et:al:2008, skardal:restrepo:2012, moreno:et:al:2004, gomezgardenes:et:al:2007}, or when the distribution of momenta is bimodal \cite{restrepo:meiss:2014}.

Both the Erd\"os-Renyi and scale-free networks satisfy the statistical equivalence property. The validity of \Asmp{1} can verified numerically. For each rotor $n$, we can estimate its effective angular velocity $\Omega_n$ by a time average, i.e., we compute 
\[
	\Omega_n  \equiv \frac{1}{T_2-T_1}[\psi_n(T_2) - \psi_n(T_1)]
\]
 for large $T_1$ and $T_2 - T_1$. As usual, $T_1$ is chosen to eliminate initial transients and $T_2$ to decrease the noise. Typical values are $T_1 = 2000$ and $T_2=10^4$. An illustration for an Erd\"os-Renyi network  is shown in Fig.~\ref{fig:OmegaDistribution}. The figure shows that the deviations of $\Omega_n$ from the average $\Omega$ are of order $0.5\%$ (and they become smaller as $T_2$ is made larger). 
Even if the rotors have the same frequency, they could have different phases. To verify this is not the case, Fig.~\ref{fig:histPsinMinusPsi1} shows a histogram of $ \psi_n(t) - \psi_1(t)$, where the histogram samples all $n \neq 1$ and at the integer times $t = 1, 2, \ldots, 2000$. 
The plot shows that the phases remain very close to each other. The tails of the distribution correspond to phase slips [i.e., $\psi_n(t) - \psi_1(t)$ rapidly changing by $2\pi$]; these slips become less frequent as the mean degree of the network is increased (not shown).


\begin{figure*}[htb]
\centering{
\subfloat{{\label{fig:ergodicAssumptionA}}\includegraphics[width=0.4\textwidth]{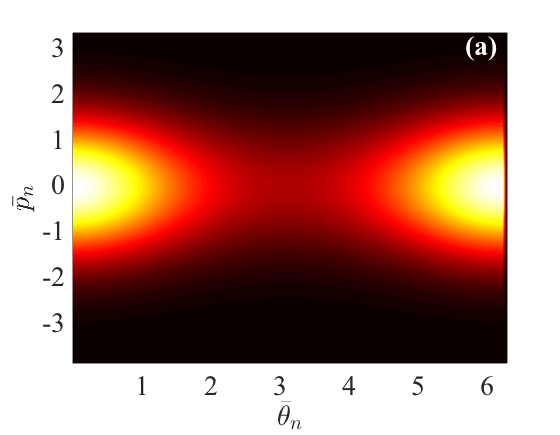}}
\subfloat{{\label{fig:ergodicAssumptionB}}\includegraphics[width=0.435\textwidth]{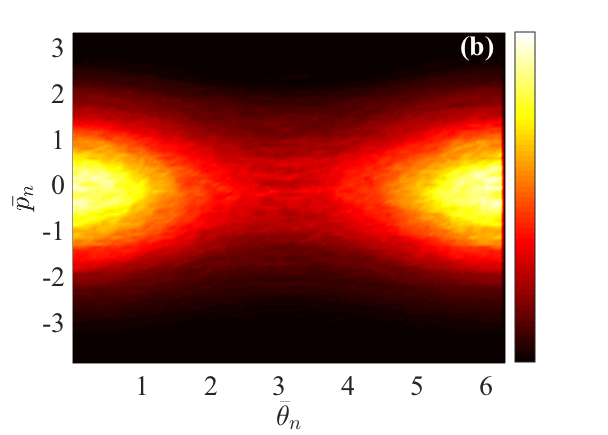}}
} 
\caption{Single oscillator phase space distribution for an Erd\"os-Renyi network ($N = 100$, $\| d \| = 10$) with $K = 1000 = 50K_c$
with an initial Gaussian distribution of momenta with mean $P_0 = 6$ and variance $\sigma_0^2 = 1$ and uniform initial phases. Panel (a) shows the theoretical Boltzmann distribution \Eq{gFunction} for node $n=1$, with  $\beta = 0.965$
and $r_1 = 0.036$. Panel (b) shows the numerical distribution of $(\bar{p}_1(t),\bar\theta_1(t))$, averaged over the time interval $\left[2000, 10000\right]$. The histograms in both panels use bins of $\Delta \theta = 0.0628$ and $\Delta p = 0.0718$.}
\label{fig:ergodicityAssumption}
\end{figure*}

Assumption \ref{assum:2} is commonly used in the analysis of the HMF model \cite{barre:et:al:2006}. The idea is that the single rotor, described by \Eq{changeThetaNBar}-\Eq{changePNBar}, is a Hamiltonian system exchanging energy with the rest of the network, which for large $N$ can be taken to be a thermal bath. This implies that, in equilibrium, the statistical behavior can be described by the Boltzmann distribution \Eq{gFunction}. The validity of this assumption is demonstrated in Fig.~\ref{fig:ergodicityAssumption}, which compares the theoretical distribution $g(\bar \theta,\bar p;r_n)$ with $r_n$ calculated from \Eq{rSigmaBesselMain} in panel (a), with a histogram of the empirical long-term distribution of the variables $(\bar \theta_n,\bar p_n)$, in panel (b) for an Erd\"os-Renyi network. The figure shows the results for the arbitrarily chosen node $n=1$---results are similar for other choices. For the phase space of the chosen rotor, the theoretical and experimental distributions are visually close. 

Assumption \ref{assum:3} would follow if the fluctuations $z_n$ were uncorrelated with $\bar\theta_n$, because then
$\langle z_n e^{-i\bar\theta_n} \rangle_t = 0$, since the fluctuations satisfy, by definition, $\langle z_n(t) \rangle_t = 0$.
If the number of connections per node is large these correlations should be weak, since $z_n$ is determined by the behavior of $\bar \theta_m$ for all the neighbors $m$ of node $n$, each of which, in turn, depends on the phases of all of their many neighbors. These heuristic arguments can be validated numerically by computing explicitly $\langle \Re(z_n e^{-i\bar\theta_n})\rangle_t$. In Fig.~\ref{fig:znIndependentOfThetaBarNAssumption} we plot the ratio $\Re \langle z_n e^{-i\bar\theta_n}\rangle_t/\| \langle r_n \cos(\bar \theta_n)\rangle_t \|$ for each of the $N=10^4$ nodes in an Erd\"os-Renyi network. For most nodes this ratio is small, less than $0.1$, though for $20$ nodes it is larger than $0.1$ and the maximum ratio is $0.16$. The validity of \Asmp{3} depends on the network average of the numerator being relatively small, and for this case we found 
\[
	 \| \Re\langle z_n e^{-i\bar\theta_n} \rangle_t \| = 1.1(10)^{-5} \ll \|\langle r_n \cos(\bar \theta_n)\rangle_t \| = 2.3(10)^{-3},
\]
confirming the assumption.

\subsection{Perturbative approximation}
\label{sec:approxOnsetSynchrony}

Equations~\Eq{rSigmaBesselMain} allow us to calculate the order parameter $R$ given a network adjacency matrix $A$, a coupling strength $K$, and the total energy. Though the numerical solutions for $r_n$ and $\sigma$ agree well with the simulations, they do not offer additional insights into how the network structure affects the general properties of the transition to synchrony. In this section, we present a perturbative analysis of \Eq{rSigmaBesselMain} near the bifurcation at $K=K_c$ that allows us to determine what properties of the network affect the value of the order parameter close to the bifurcation. 

In order to do this, we will solve this system perturbatively, assuming that $\Delta K = K-K_c \ll 1$---the coupling constant is just beyond the critical value. We solve the system \Eq{rSigmaBesselMain}, i.e.,
\begin{align*} 
r_n &= \frac{1}{N} \sum_{m=1}^{N} A_{nm} v\left(\frac{K r_m}{\mu}\right) \enspace, \\
\mu& = \sigma_0^2 + \frac{K}{N} \sum_{n=1}^{N} r_n v\left(\frac{K r_m}{\mu}\right) \enspace,\end{align*}
using $\mu = \sigma^2$ for the variance and setting the initial potential energy $U_0$ to zero, e.g., for initial conditions in the incoherent state in the limit $N \to \infty$. Introducing a formal small parameter $\eps$, the perturbative expansion takes the form
\begin{equation}\label{eq:seriesExpansions}
\begin{split}
	K &= K_c + \eps \Delta K,\\
	r_n &= \eps^{1/2} r_n^{(1)} + \eps r_n^{(2)} + \eps^{3/2} r_n^{(3)} + \mathcal{O}(\eps^2),\\
\mu &=  \mu^{(0)} + \eps\mu^{(1)} +  \mathcal{O}(\eps^{3/2}),
\end{split}
\end{equation}
where we have anticipated already that $r_n \sim (\Delta K)^{1/2}$ and have included only terms up to the order necessary to determine $r_n^{(1)}$ in the analysis that follows. Inserting these in \Eq{rSigmaBesselMain} and expanding in powers of $\eps$ we obtain
at zeroth order,
\[
	\mu^{(0)} = \sigma_0^2,
\]
as expected. The next terms, of order $\eps^{1/2}$, imply
\[
		\left(I - \frac{K_c}{2\sigma_0^2 N} A\right) \br^{(1)} = 0,
\]
which gives
\begin{equation}\label{eq:kkc}
\begin{split}
	K_c 	  &= \frac{2\sigma_0^2 N}{\lambda}, \\
	\br^{(1)} &= C \bu. 
\end{split}
\end{equation}
Here $\bu$ and $\lambda$ are the principal eigenvector and eigenvalue of $A$, and $C$ is a normalization constant to be determined (as we will see, the product $C\bu$ does not depend on the normalization of $\bu$). This result is in agreement with the linear stability calculation of \Sec{linearStabilityAnalysis} [cf. \Eq{criticalKFirstOrder}]. The terms of order $\eps^1$ lead to $\br^{(2)} \propto \bu$ (although this will not be used), and to
\begin{align}\label{eq:mu1}
\mu^{(1)} = \frac{K_c^2 C^2}{2\sigma_0^2} \| \bu^2 \| \enspace.
\end{align}
Here $\bu^k$ denotes the vector with components $u_n^k$, and the $\|...\|$ the network average \Eq{NodeAverage}, as usual. Finally, the terms of order $\eps^{3/2}$ yield
\begin{equation}\label{premu}
\begin{split}
    &\left(I - \frac{K_c}{2\sigma_0^2 N} A\right) \br^{(3)} = \\
    	&\frac{C}{16 \sigma_0^6 N}  A
	\left [ 8\sigma_0^2 (\Delta K \sigma_0^2 - K_c \mu^{(1)}) \bu -C^2 K_c^3 \bu^3\right].
\end{split}
\end{equation}
In order to eliminate the unknown vector $\br ^{(3)}$, we multiply (\ref{premu}) on the left by $\bu^T$. Since $A^T = A$, then $\bu^T A = \lambda \bu^T$ and using \Eq{kkc}, the left hand side vanishes, giving the solvability condition
\[
	0 =  8 \sigma_0^2 (\Delta K \sigma_0^2-K_c \mu^{(1)}) \| \bu^2\| -C^2 K_c^3 \| \bu^4\|.
\]
Using \Eq{mu1}, this determines $C$:
\[
	C = \left( \frac{8\sigma_0^4 \|\bu^2 \|}
		{\| \bu^4 \| + 4 \| \bu^2 \|^2 } 
		\frac{\Delta K}{K_c^3}\right)^{1/2} \enspace .
\]
Finally, using the definition \Eq{globalOrderParam},
\[
 	R = \frac{1}{\| d \|}\sum_{n=1} r_n = \eps^{1/2} C N \frac{\| \bu \|}{\| d \|} + \mathcal{O}(\eps) 
\]
 and \Eq{kkc}, we find (dropping the formal parameter $\eps$), the main result of this section,
\begin{align}\label{eq:globalOrderRInDeltaK}
	R \approx  G \sqrt{\frac{\Delta K}{K_c } } ,
	  \quad G \equiv 
		\frac{\lambda}{\| d \|} 
		\left({\frac{2 \| \bu^2 \| \| \bu \|^2}
		{\| \bu^4 \| + 4 \| \bu^2 \|^2 } }\right)^{1/2}. 
\end{align}

This expression provides some insight into the effect of heterogeneity on synchronization through the factor $G$. For example, for an uncorrelated network for which $u_n \propto d_n$ \cite{restrepo:ott:hunt:2007}, $G = \sqrt{2/5}$ for a regular, homogeneous graph with $d_n = d$, while $G \to 0$ for when the degree distribution is heterogeneous so that $\| d^4 \| \to \infty$ in the limit $N\to \infty$. Thus, in this case we find that heterogeneity tends to make the transition to synchrony less sharp.

To illustrate this, compare this theoretical result to the numerical results for the time averaged order parameter $\langle R\rangle_t$ in Fig.~\ref{fig:R_vs_K_theoretical} for an Erd\"os-Renyi (homogeneous) and a scale-free (heterogeneous) network with the same size and mean degree. The dashed lines show the approximation \Eq{globalOrderRInDeltaK}. We find $G=0.633$ for the Erd\"os-Renyi network, larger than $G=0.354$ for the scale-free network, as we would expect.
The second-order approximation agrees with the numerical results for the scale-free network whenever $K > K_c$; however, $R$ for the Erd\"os-Renyi network is not well approximated for higher values of $\Delta K$. Of course, $\Delta K$ is assumed to be small in the derivation above, so there is no reason for agreement for large $\Delta K$.

\subsection{Large $K$ limit}
\label{sec:largeK}

Figures \ref{fig:ERstdP} and \ref{fig:Assumptions1} suggest that $R$ tends to an asymptotic value, $R \to \hat R < 1$, as $K \to \infty$. In this section we will study this limit and explore how $\hat R$ depends on the network. To begin our analysis, we divide \Eq{rSigmaBesselMain} by $K$ and let $\eta \equiv \sigma^2/K$ to obtain
\begin{align*}
r_n &= \frac{1}{N} \sum_{m=1}^{N} A_{nm} v\left(\frac{r_m}{\eta}\right) \enspace ,\\
\eta & = \frac{\sigma_0^2+ 2U_0/N}{K} + \frac{1}{N} \sum_{n=1}^{N} r_n v\left(\frac{r_n}{\eta}\right) \enspace.
\end{align*}
Under the hypothesis that $\lim_{K \to \infty} \eta =  \hat\eta$ is finite, in the limit $K \to \infty$ the system above reduces to 
\begin{equation}\label{eq:KtoInfinity}
\begin{split}
\hat r_n &= \frac{1}{N} \sum_{m=1}^{N} A_{nm} v\left(\frac{\hat r_m}{\hat\eta}\right) \enspace, \\
\hat\eta & = \frac{1}{N} \sum_{n=1}^{N} \hat r_n v\left(\frac{\hat r_n}{\hat\eta}\right) \enspace ,  
\end{split}
\end{equation}
for the asymptotic values $\hat r_n$, and $\hat \eta$.
A solution can be found numerically as the fixed point of the relaxed iteration scheme
\begin{align*}
\hat r_n^{t+1} &= (1-\beta)\frac{1}{N} \sum_{m=1}^{N} A_{nm} v\left(\frac{\hat r_m^{t}}{\hat\eta^{t}}\right) + \beta \hat r_n^{t} \enspace,\\
\hat\eta^{t+1}  & = \frac{1}{N} \sum_{n=1}^{N} \hat r_n^{t}  v\left(\frac{\hat r_n^t}{\hat\eta^{t}}\right) \enspace,
\end{align*}
where $0 \leq\beta < 1$ is a relaxation factor included to obtain convergence to a fixed point.

\begin{figure}[t]
\centering{
\includegraphics[width=0.4\textwidth]{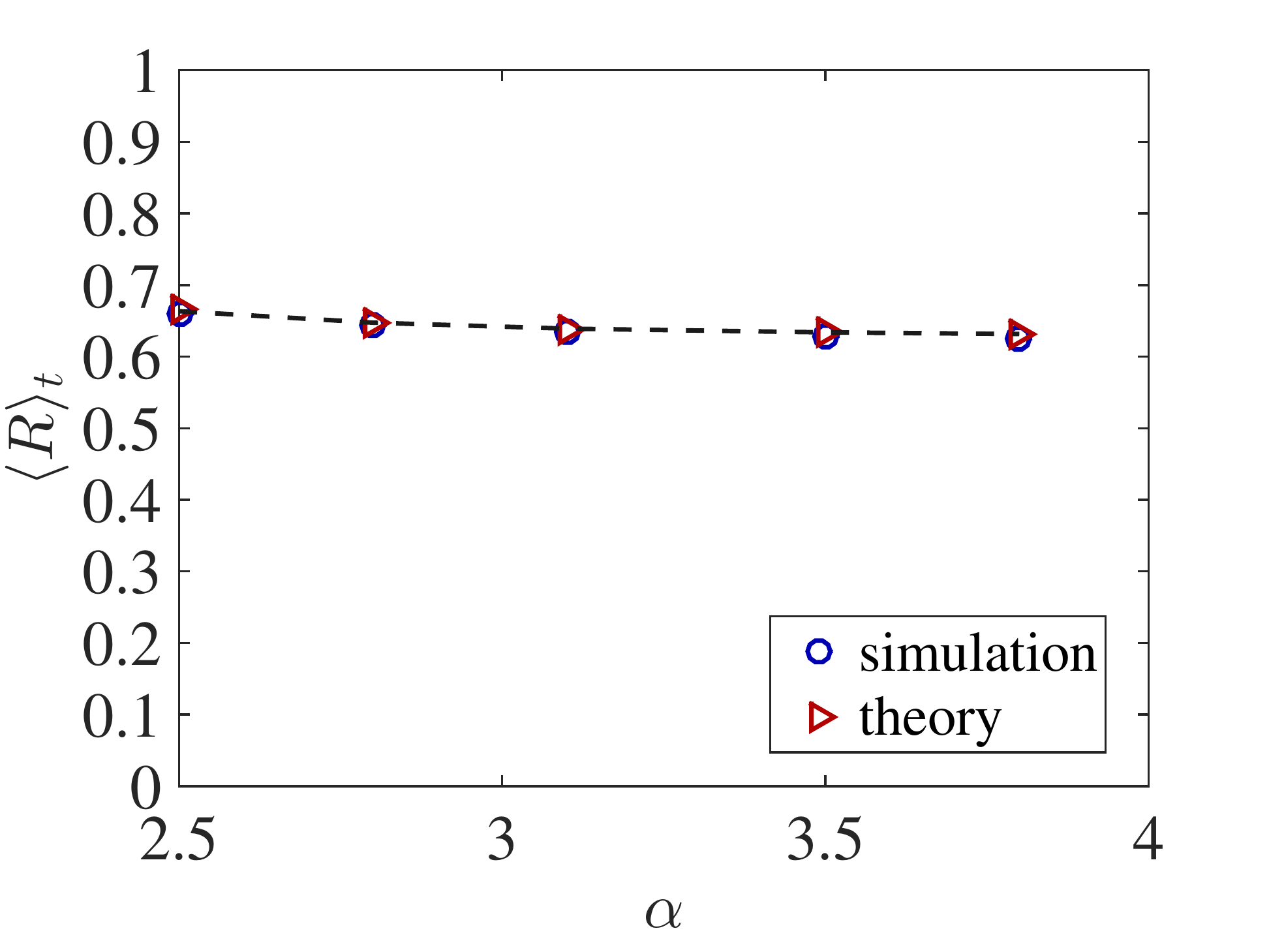}
\caption{Asymptotic order parameter, $\hat R$, for scale-free networks with varying degree exponent $\alpha$. The red triangles show the iterative solution of \Eq{KtoInfinity}, and the blue circles show $\langle R \rangle_t$ from simulations using a large coupling constant, $K=10^4$. The dashed line shows the approximation from \Eq{rd}.}
\label{fig:maxSynchrony}
}
\end{figure}

Figure~\ref{fig:maxSynchrony} shows the global order parameter $R$ obtained from both the numerical solution of \Eq{KtoInfinity} (red triangles) and the numerical solution of the full system, \Eq{dPhase_dT}-\Eq{dMomentum_dT} (blue circles) for scale-free networks as a function of the exponent $\alpha$ of the degree distribution \Eq{powerlaw}. For the simulations, $K = 10^4$ and the observed order parameters are about $1\%$ smaller than the theory, which assumes $N \to \infty$ This difference is not visible on the scale of the figure. Remarkably, the asymptotic value of the order parameter is nearly independent of the exponent $\alpha$.

An additional approximation can be made, following \cite{ichinomiya2004frequency}, if we assume that the local order parameters are proportional to the nodal degrees, $\hat r_n = B d_n$, where $B$ is a constant to be determined. This approximation works well for homogeneous networks without correlations \cite{restrepo:ott:hunt:2007}; for scale-free networks, it works well for power law exponents $\alpha > 3$ \cite{restrepo:ott:hunt:2005}. If we replace $\hat r_n$ by $B d_n$ in \Eq{KtoInfinity} and sum the first equation over $n$ we obtain, using $d_m = \sum_{n} A_{nm}$, that
\begin{align}
B N \| d \| &= \frac{1}{N} \sum_{m=1}^{N} d_m v\left(\frac{ B d_m}{\eta}\right) \enspace, \label{eq:etaBessel_added}\\
\eta  & = \frac{B}{N} \sum_{n=1}^{N} d_n v\left(\frac{B d_n}{\eta}\right) \enspace.  \label{eq:sigmaSqBessel_added}
\end{align}
Comparing the two equations we see that $\eta = B^2 N \| d \|$. From the definition of the global order parameter we find $\hat R = \sum_{n} \hat r_n /\| d \| = B \sum_{n} d_n /\| d \| = N B$. Using these two in \Eq{etaBessel_added} we find a nonlinear equation for the single variable $\hat R$
\begin{align}\label{eq:rd}
\hat R \| d \| &= \frac{1}{N} \sum_{m=1}^{N} d_m v\left(\frac{ d_m}{\hat R \| d \|}\right) \enspace.
\end{align}
This equation can be solved numerically using standard root-finding tools, and produces the dashed line shown in Fig.~\ref{fig:maxSynchrony}.

If the degrees of individual nodes are not known, but the degree distribution $P(k)$ is known, one can approximate the sum in \Eq{rd} by an integral to obtain an implicit equation for $\hat R$ (using the dummy variable $k$ instead of $d$)
\begin{align*}
\hat R \| d \| &= \int k P(k) v\left(\frac{k}{\hat R \| d \|}\right)\text{d}k \enspace. 
\end{align*}
 Finally we note that, although this was not pursued here, a similar mean mean field approach (i.e., $r_n = B d_n$) could be used to further study the system \Eq{rSigmaBesselMain}.

\section{Conclusion}
\label{sec:conclusion}

In this paper we studied the HMF model where the interactions between rotors are described by a weighted adjacency matrix. We found that, as in other dynamical systems on networks (e.g., \cite{restrepo:ott:hunt:2005}, \cite{mieghem:2012}, \cite{larremore:et:al:2012}), the transition to synchrony occurs at a value of the coupling constant inversely proportional to the largest eigenvalue, $\lambda$, of the adjacency matrix. Thus the primary effect of network structure on this aspect of the dynamics is $\lambda$. 

We obtained a set of equations that determine the set of local order parameters in the synchronized state. These equations relied on three assumptions that were verified a posteriori for the Erd\:os-Renyi and scale free networks studied in Sec.~\ref{sec:synchronizedState}. Of these assumptions, the most important is that the network is constructed in such a way that the neighbors of all nodes share the same statistical properties. This assumption is not satisfied, for example, by networks with strong community structure. While this seems restrictive, the class of networks to which our results apply include networks with heterogeneous degree distributions (e.g., scale-free networks) and networks with degree-degree correlations. It is also expected that some of our results could be extended to networks with community structure.

Our main result is a method to quantitatively explore the effect of network heterogeneity on the transition to synchrony, resulting in \Eq{globalOrderRInDeltaK}. In addition to determining $K_c$, the critical coupling constant, network heterogeneity also affects the sharpness of the transition, with more heterogeneous networks having a less pronounced transition. However, even though such heterogeneity (as represented the degree distribution) has a strong effect both on the location and sharpness of the onset of synchrony, it seems to have little effect on the degree of synchronization for large coupling. 

In conclusion, our results show that many of the phenomena that have been observed for the all-to-all coupled HMF model persist for more complex networks, and that the onset of synchrony is determined by spectral properties of the coupling matrix.

\acknowledgments
JDM was partially supported by NSF grant DMS-1211350.

\appendix

\section{Derivation of the dispersion relation}\label{dispersion}

In this Appendix we derive the dispersion relation \Eq{dispersion} by studying the evolution of the perturbations ($\delta \theta_n, \delta p_n)$ to the incoherent initial state $(\bar \theta_n(t), \bar p_n(t)) = (\bar p_n t + \theta_n^0, \bar p_n^0)$, where  $\theta_n^0$ are uniformly distributed in $[0,2\pi)$, and the initial momenta $\bar p_n^0$ are arbitrary. 
Integrating the perturbed ODEs \Eq{linearizedODEs} formally with respect to time gives
\begin{align*} 
        \delta \theta_n(t) = &\int_{t_0}^t \delta p_n(t')dt' +  \delta \theta_n(t_0) \enspace,\\
        \delta p_n(t) = &\frac{K}{N} \sum_{m=1}^{N} A_{nm} \int_{t_0}^t \cos[\bar\theta_m(t')-\bar\theta_n(t')] \delta \theta_m(t')dt' \\
        & + \delta p_n(t_0) \enspace,
\end{align*}
Defining $C_n(t) \equiv \delta \theta_n(t) -  \delta \theta_n(t_0)$ 
and integrating the second equation from $t_0$ to $t$, we find
\begin{equation}
\begin{split} \label{eq:withI}
 C_n(t) = &\frac{K}{2N}  \int_{t_0}^t \int_{t_0}^{t'} e^{-i\theta_n(t'')}
			\sum_{m=1}^{N} A_{nm}[ e^{i\theta_m(t'')} \\ 
	    &+ e^{2i\theta_n(t'')}e^{-i\theta_m(t'')}] C_m(t'') dt'' dt' + I \enspace,
\end{split}
\end{equation}
where $I$ is
\begin{align*}
	I = &\frac{K}{N} \sum_{m=1}^{N} A_{nm} \int_{t_0}^t \int_{t_0}^{t'} 
		\cos\left[\bar\theta_m(t'')-\bar\theta_n(t'')\right] \delta\theta_m(t_0)dt'' dt' \\
	    &+ (t-t_0)\delta p_n(t_0) \enspace.
\end{align*}
Defining $B_k(t) \equiv \sum_{m=1}^N A_{km}e^{i\theta_m(t)}C_m(t)$, multiplying equation \Eq{withI} by $A_{kn} e^{i\bar\theta_n(t)}$ and summing over $n$ yields
\begin{align*}
	B_k(t) &= \frac{K}{2N} \sum_{n=1}^N A_{kn} \int_{t_0}^t \int_{t_0}^{t'}
		    e^{i(\bar\theta_n(t) - \bar\theta_n(t''))} [B_n(t'')  \\
		& +e^{2i\bar\theta_n(t'')} B^*_n(t'')]dt''dt'  + 
			\sum_{n=1}^N A_{kn} e^{i\theta_n(t)} I.
\end{align*}
To find the dispersion relation, we assume exponential growth of the perturbations, i.e., $B_k(t) = b_k e^{s t}$, where $\Re(s) > 0$. Using this, we get that
\begin{equation}
\begin{split}\label{eq:integrals}
	b_k e^{s t} =& \frac{K}{2N} \sum_{n=1}^N A_{kn} \int_{t_0}^t \int_{t_0}^{t'} 
		e^{i(\bar\theta_n(t) - \bar\theta_n(t''))}[b_ne^{s t''} \\ 
	&+e^{2i\bar\theta_n(t'')} b_n^* e^{s^* t''}]dt''dt' 
		+ \sum_{n=1}^N A_{kn} e^{i\theta_n(t)} I.
\end{split}
\end{equation}
Since we are assuming $\Re(s) > 0$, the left hand side of \Eq{integrals} grows exponentially with $t$. However, the term $I$ grows at most quadratically,
\[
	|I| \leq  \tfrac12 (t-t_0)^2\frac{K}{N} \sum_{m=1}^{N} A_{nm}  \delta\theta_m(t_0) 
		 + (t-t_0)\delta p_n(t_0),
\]
and therefore as $t\to \infty$ the first term on the right-hand side of \Eq{integrals} must balance the left hand side. Replacing $\bar \theta_n = \bar p_n t + \theta_n(0)$, we obtain
\begin{align}\label{withoutI}
b_k = &\frac{K}{2N} \sum_{n=1}^N A_{kn} b_n \int_{t_0}^t \int_{t_0}^{t'} e^{(s - i \bar p_n)(t''-t)}dt''dt'  \\
& + \frac{K}{2N} \sum_{n=1}^N A_{kn} e^{2i \theta_n(0)} \int_{t_0}^t \int_{t_0}^{t'}  b_n^* e^{(s^*+i\bar p_n) t'' - (s - i \bar p_n)t}dt''dt'.
\end{align}
Since the angles $\theta_n(0)$ are uniformly distributed in $[0,2\pi)$, the second term can be neglected for large $N$. Integrating the first term and taking the limit $t \to \infty$ with $\Re(s) > 0$ we finally obtain \Eq{dispersion}.


\bibliographystyle{unsrt}
\bibliography{hmf_network}

\end{document}